\documentclass[sigconf,nonacm]{acmart}

\usepackage{makecell}
\usepackage{tabularx}
\usepackage{subcaption}
\usepackage{siunitx}
\usepackage{cleveref}
\usepackage{hyperref}
\usepackage{minted}
\usepackage{placeins}
\usepackage{xcolor}
\usepackage{algorithm}
\usepackage{algpseudocode}
\algrenewcommand\textproc[1]{\textsf{#1}}
\usepackage{booktabs}
\usepackage{longtable}
\usepackage{multirow}
\algrenewcommand\algorithmiccomment[1]{\hfill{\color{gray}\(\triangleright\) #1}}

\definecolor{commentblue}{RGB}{70,100,140}
\algrenewcommand\algorithmiccomment[1]{\hfill{\color{commentblue}\(\triangleright\) #1}}

\newcommand{\sysname}{\textsc{AEGIS}}

\usepackage{tikz}
\DeclareRobustCommand{\stepnum}[1]{%
  \tikz[baseline=(char.base)]{
    \node[
      shape=circle,
      fill=black,
      text=white,
      inner sep=0.3pt,
      minimum size=1em
    ] (char) {\small\bfseries #1};
  }%
}


\begin{document}

\title{\sysname{}: Runtime-Guided GPU Collocation for Multi-Tenant Deep Learning Training}

\author{Ehsan Yousefzadeh-Asl-Miandoab}
\affiliation{
  \institution{IT University of Copenhagen}
  \country{Denmark}
}

\author{Büşra Karatay Demiray}
\affiliation{
  \institution{HES-SO University of Applied Sciences and Arts}
  \country{Switzerland}
}

\author{Florina M. Ciorba}
\affiliation{
  \institution{University of Basel}
  \country{Switzerland}
}

\author{Pamela Delgado}
\affiliation{
  \institution{HES-SO University of Applied Sciences and Arts}
  \country{Switzerland}
}

\author{Pınar Tözün}
\affiliation{
  \institution{IT University of Copenhagen}
  \country{Denmark}
}

\renewcommand{\shortauthors}{Ehsan Yousefzadeh-Asl-Miandoab et al.}

\begin{abstract}
Deep learning training commonly runs on shared multi-tenant GPU servers, where exclusive allocation provides isolation but can leave GPU resources underutilized and increase queueing time. Collocating compatible workloads can improve efficiency, but interference-agnostic placement can cause severe slowdowns, while inaccurate or unavailable memory information can lead to out-of-memory (OOM) failures. A scheduler must therefore consider both memory feasibility and whether a GPU's ongoing activity permits additional work.

We present \sysname{}, a server-scale runtime scheduling system for controlled collocation of deep learning training workloads on shared multi-GPU servers. \sysname{} coordinates memory feasibility, mandatory post-placement observation, runtime-pressure filtering, placement, and OOM-aware recovery within one scheduling loop. After each placement, it temporarily withholds the affected GPU(s) from further collocation until the newly launched workload shows GPU activity and a short activity-anchored monitoring window is complete. It then derives pressure indicators from low-overhead telemetry, updates runtime-pressure eligibility, and applies the configured placement policy using the relevant maintained GPU state. OOM-aware recovery retries failed jobs under progressively safer memory conditions and falls back to exclusive execution when needed. This lightweight online control loop does not require costly offline pairwise compatibility profiling.

We evaluate \sysname{} using vision, Transformer, recommendation, and LLM-style training workloads across three production-derived workload traces. \sysname{} reduces geometric-mean makespan by 16\% relative to Lucid~\cite{lucid_asplos}, 21\% relative to Horus~\cite{horus}, and 27\% relative to exclusive allocation. Sensitivity studies show that activity-anchored observation and runtime-pressure filtering balance overly conservative isolation against interference-agnostic collocation, improving makespan while limiting sharing-induced per-task slowdown.
\end{abstract}

\maketitle

\section{Introduction}
\label{sec:intro}

Shared GPU servers and clusters are now a common substrate for deep learning training, with workloads from multiple users competing for a limited pool of accelerators. Exclusive execution provides strong isolation, but it can leave compute, memory-bandwidth, or device-memory capacity unused while other jobs wait for resources~\cite{MLaaS,AnalysisOfMultiTenantGPUClusters,hu2021characterization,10.1145/3638757,gao2022deep}. Collocation can improve server efficiency by allowing compatible workloads to share this residual capacity.

Modern GPUs provide mechanisms such as time slicing~\cite{gpu_stream}, Multi-Process Service (MPS)~\cite{nvidia_mps}, and hardware partitioning~\cite{mig-main-guide-pdf}. These mechanisms expose different sharing and isolation modes~\cite{robroek2023analysis, CaoSIAK23}. However, they do not determine when another job should be placed, which GPU(s) should host it, or how the scheduler should respond when a placement fails at runtime. Effective collocation therefore remains a runtime scheduling problem.

Collocation decisions are difficult because GPU compatibility is multidimensional. Two workloads may fit together in device memory yet contend for execution resources, resident-warp capacity, or memory bandwidth. Conversely, workloads with similar memory footprints can interact differently depending on their active-execution, occupancy, and memory-system behavior. Memory availability therefore determines whether a placement is feasible, but it does not determine the interference cost of sharing. In addition, inaccurate peak-memory estimates or incomplete profiling information can turn an apparently feasible placement into an out-of-memory (OOM) failure.

\begin{figure}[t]
\centering
\includegraphics[width=0.95\columnwidth]{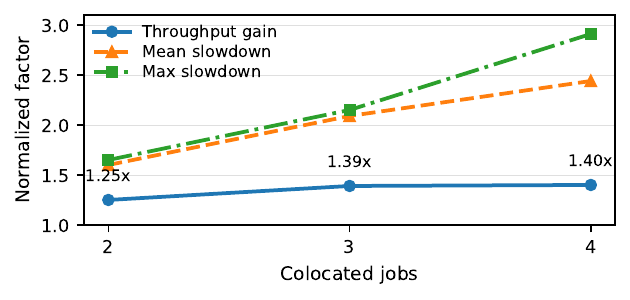}
\caption{
Interference-agnostic MPS-enabled collocation on an A100 40\,GB GPU. We start with ResNet50 (BS128) and VGG16 (BS64), then add Xception (BS64), followed by ResNet50 (BS32), all on ImageNet~\cite{imagenet}, where BS denotes batch size. All jobs complete in every configuration. Throughput gain is the sum of the jobs' solo runtimes divided by the collocated makespan.
}
\label{fig:motivation-interference-agnostic-collocation}
\end{figure}

\Cref{fig:motivation-interference-agnostic-collocation} illustrates the throughput--slowdown trade-off under interference-agnostic MPS collocation. Beyond three jobs, additional memory-feasible collocation yields little further throughput gain relative to serial solo execution, while substantially increasing mean and worst-case slowdown. This motivates treating memory feasibility and interference risk as separate scheduling decisions. The key challenge is therefore not whether workloads can share a GPU, but when they should share one and under what runtime conditions.

Existing systems improve GPU sharing in different ways, but they make different assumptions about what information and control are available. Gandiva combines time slicing, migration, introspective scheduling, and opportunistic collocation~\cite{Gandiva,gandiva-fair}; Lucid uses offline profiling and interpretable models to guide workload packing~\cite{lucid_asplos}; Horus estimates workload GPU-memory demand and predicts GPU utilization as an interference proxy for placement~\cite{horus}; and Orion intercepts and schedules GPU operations according to their compute and memory requirements on a single GPU~\cite{strati2024orion}. These systems expose useful design points, but they require different combinations of offline profiling, representative prediction data, workload instrumentation, or fine-grained execution control.

This leaves an important gap for shared multi-GPU training servers. A scheduler may need to place previously unseen workloads without profiling every workload pair or relying entirely on estimators whose accuracy can vary across workloads, configurations, and GPU platforms~\cite{yousefzadehGPUMemoryEstimation}. Free memory alone cannot reveal whether a running workload is already placing substantial pressure on compute or memory-system resources, while telemetry collected immediately after launch may capture initialization rather than meaningful GPU execution. Consequently, placement decisions made immediately after launch may rely on stale or unrepresentative GPU state. Because imperfect memory information can still cause OOM failures, these constraints motivate a runtime control loop that separates memory feasibility from interference risk, withholds further collocation until meaningful activity is observed, refreshes runtime state, and recovers from failed placements.

To address this gap, we present \sysname{}, a server-scale runtime scheduling system for controlled collocation of deep learning training workloads. \sysname{} processes queued jobs, enforces each job's GPU-count requirement, and applies a configurable placement policy after mandatory post-placement observation. If no suitable placement exists, the job remains queued.

\sysname{} combines mandatory post-placement observation with runtime-pressure filtering. After each placement, it temporarily withholds the affected GPU(s) from further collocation, anchors monitoring at the newly launched workload's first observed GPU activity, and waits for a complete activity-anchored window. The completed window refreshes available memory and pressure indicators, after which configured thresholds establish runtime-pressure eligibility. Depending on the selected policy, placement decisions may then use memory feasibility and runtime-pressure eligibility to select among the remaining candidate GPUs.

\sysname{} derives runtime-pressure eligibility from activity-anchored monitoring rather than relying only on prediction or exhaustive profiling. For memory-aware policies, memory capacity remains a hard feasibility constraint, while eligibility determines whether an otherwise feasible GPU is currently suitable for additional work. Because memory information can be imperfect, \sysname{} complements this preventive filtering with a recovery path: OOM-affected workloads are typically detected early during initialization or warm-up, requeued with higher priority, retried under progressively safer memory conditions, and ultimately run exclusively if needed. Together, these mechanisms form an online control loop for queue-level GPU sharing based on runtime eligibility and OOM-aware recovery.

Our contributions are:
\begin{list}{\labelitemi}{\leftmargin=1.5em}

\item We present \textbf{\textit{\sysname{}}}, a modular server-scale runtime scheduler integrating configurable memory inputs, mandatory post-placement observation, runtime-pressure filtering, placement policies, and recovery.

\item We introduce \textbf{\textit{activity-anchored post-placement observation}}, which temporarily withholds the affected GPU(s) from further collocation until the newly launched workload shows GPU activity and a complete 30\,s monitoring window is collected. The resulting telemetry refreshes available memory and establishes \textbf{\textit{runtime-pressure eligibility}} for subsequent placement decisions.

\item We design \textbf{\textit{OOM-aware recovery}} that uses failed attempts to retry jobs under progressively safer memory conditions and falls back to exclusive execution when needed.

\item We evaluate memory input, placement policy, runtime-pressure eligibility, monitoring-window and threshold sensitivity, arrival order, and limited-telemetry operation.
Across three workload traces, \sysname{} reduces makespan by 16\%, 21\%, and 27\% relative to Lucid~\cite{lucid_asplos}, Horus~\cite{horus}, and exclusive allocation, respectively.
\end{list}

\section{Related Work}
\label{sec:related}

\paragraph{Schedulers for deep learning training.}
Several systems build scheduling abstractions on top of GPU execution mechanisms. Gandiva exploits the iterative structure of training to support time slicing, migration, and adaptive scheduling~\cite{Gandiva}. Salus introduces fast job switching and dynamic memory sharing through iteration-level scheduling~\cite{SalusFineGrainGPUSharingDL}. PipeSwitch pipelines model-state movement to provide low-overhead context switching between deep learning applications~\cite{PipeSwitch}. AntMan co-designs the scheduler and training framework to dynamically scale GPU memory and computation allocations~\cite{AntMan}. These systems primarily provide mechanisms for multiplexing or reallocating GPU resources, whereas \sysname{} focuses on maintaining trustworthy per-GPU runtime state and deciding whether an active GPU should receive additional work.

\paragraph{Fairness and elastic scheduling.}
Elastic schedulers adapt the number or type of resources assigned to training jobs during execution~\cite{optimus,AFS/CoDDL,KungFu,Pollux,sia,lyra,elasticFlow}. Pollux, for example, jointly adapts resource allocations and training configurations to optimize cluster-wide goodput~\cite{Pollux}. Other systems optimize cluster-level objectives such as job completion time, makespan, placement efficiency, or fairness~\cite{Tiresias,Themis,gandiva-fair,HiveD,gavel,shockWave}. Themis introduces finish-time fairness for long-running, placement-sensitive training jobs~\cite{Themis}, while Gavel generalizes scheduling policies across heterogeneous accelerators and explicitly models space sharing~\cite{gavel}. These cluster-level objectives are complementary to \sysname{}, which uses FIFO queue ordering and focuses on runtime-guided collocation within a multi-GPU server rather than fairness or elastic resource reconfiguration.

\paragraph{Profile-guided and prediction-based collocation.}
The closest related work uses workload information to predict whether jobs can beneficially share a GPU. Horus predicts GPU utilization from computation-graph features and treats predicted utilization as a proxy for interference when selecting placements~\cite{horus}. This avoids online profiling of each submitted job, but its decisions depend on the accuracy and portability of the learned utilization model. \sysname{} instead observes the current pressure of already-running GPUs and separates this runtime eligibility signal from the memory-information source used to establish placement feasibility.

Lucid profiles workloads both individually and under pairwise collocation to characterize sharing behavior, then trains an interpretable classifier that assigns sharing classes from workload features~\cite{lucid_asplos}. At scheduling time, an unseen workload is profiled individually for a bounded period, and the resulting features are used to predict its sharing class and guide packing. This avoids profiling every new workload pair, but still requires a representative offline training set and per-workload profiling before scheduling. \sysname{} instead derives runtime-pressure eligibility from activity-anchored monitoring after placement, allowing operation without standalone workload profiling.

Orion targets finer-grained GPU sharing by transparently intercepting kernel launches and scheduling individual operators according to their compute and memory requirements~\cite{strati2024orion}. Its operator-level control operates at much finer timescales than job placement but requires GPU-call interception and operator characterization. Muri instead exploits temporal complementarity across CPU, GPU, network, and storage activity through multi-resource interleaving~\cite{Muri}. Both systems operate at granularities different from \sysname{}'s server-scale job placement.

Unlike approaches centered on offline profiling or prediction, \sysname{} combines lightweight runtime eligibility filtering with recovery when imperfect memory information leads to OOM failures.

\section{\sysname{} Design}
\label{sec:design}

\sysname{} is a server-scale runtime scheduling system for controlled collocation of deep learning training workloads. It coordinates memory feasibility, mandatory post-placement observation, runtime-pressure filtering, placement policy, and OOM-aware recovery to improve GPU efficiency.

\subsection{System Overview}
\label{sec:design:overview}

\begin{figure}[t]
\centering
\includegraphics[width=1\linewidth, trim={0 0 0 0.1cm},clip]{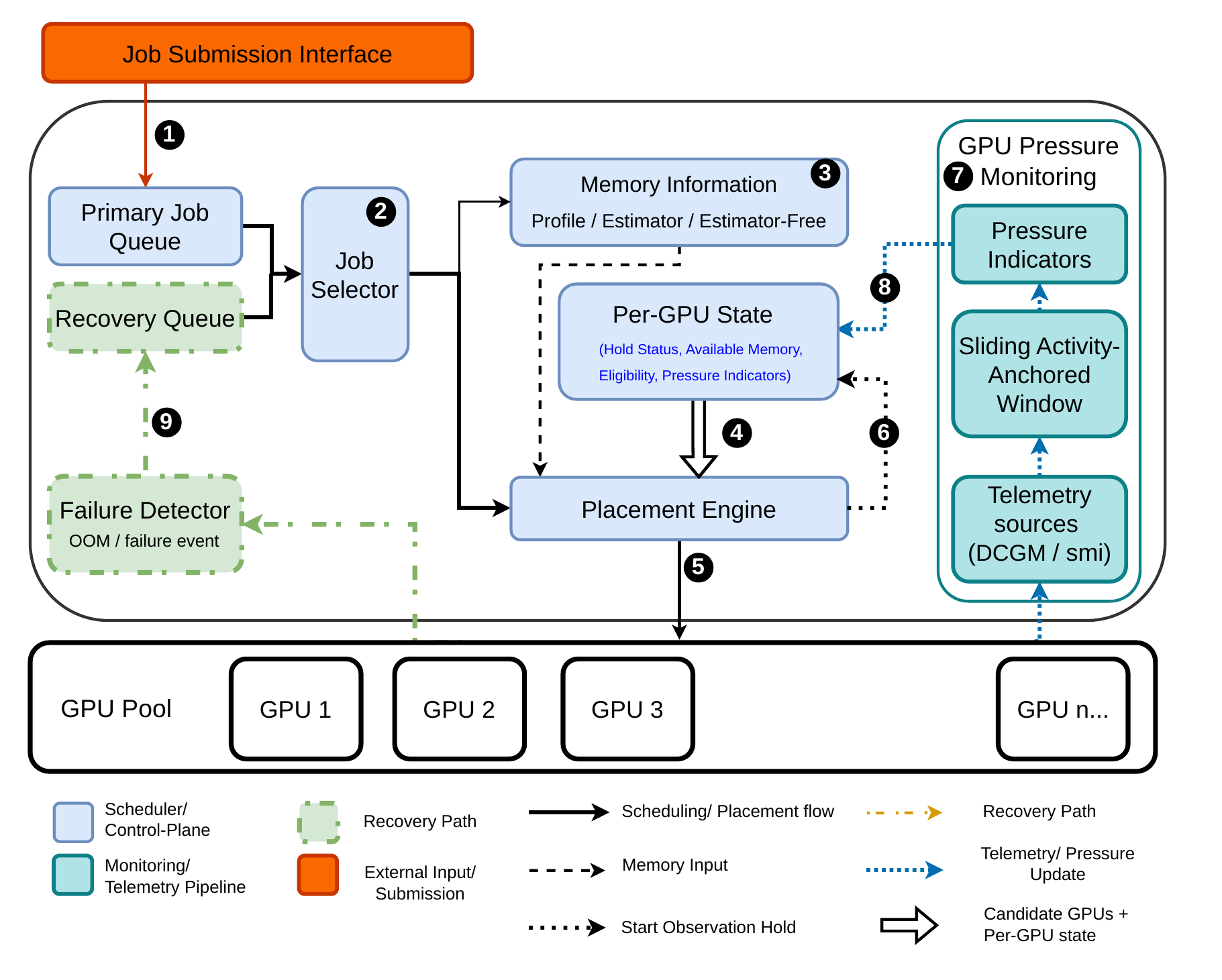}
\caption{Overview of \sysname{}. Numbered markers show the control flow: submit a job (\stepnum{1}), select the next job (\stepnum{2}), read its memory input (\stepnum{3}), read the current per-GPU state (\stepnum{4}), place the job (\stepnum{5}), start a mandatory post-placement observation hold (\stepnum{6}), detect activity and update the activity-anchored monitoring window (\stepnum{7}), release the hold and refresh runtime-pressure eligibility (\stepnum{8}), and recover from an OOM (\stepnum{9}).}
\label{fig:carma_overview}
\end{figure}

\Cref{fig:carma_overview} shows \sysname{}'s architecture, and \Cref{alg:aegis-scheduler} summarizes its online control flow. Submitted jobs enter the primary queue (\stepnum{1}), while OOM-affected jobs enter the higher-priority recovery queue. The job selector chooses the next job from either queue (\stepnum{2}) and passes it to the placement engine.

The placement engine reads the job's memory input (\stepnum{3}) and current per-GPU state (\stepnum{4}), including observation-hold status, latest available memory, runtime-pressure eligibility, and pressure metadata. The configured policy determines which of these inputs guide candidate construction and selection. The placement engine first excludes GPUs under post-placement observation hold, then builds the candidate set using the memory and runtime-pressure information required by the configured policy. It may prefer an all-idle placement when one is available, avoiding unnecessary interference when idle capacity already satisfies the job's requirements. If enough candidates remain to satisfy the job's GPU-count requirement, the policy ranks them and the job is launched on the selected GPU(s)~(\stepnum{5}).

After placement, the affected GPU(s) enter mandatory observation hold and are excluded from further collocation (\stepnum{6}). The monitoring pipeline detects the newly launched workload's first observed GPU activity, fills an activity-anchored window, and updates available memory and pressure indicators (\stepnum{7}). Once the initial window is complete, the hold is released and the saturation rule establishes runtime-pressure eligibility (\stepnum{8}). Subsequent telemetry updates maintain the latest available-memory value and refresh the pressure indicators and eligibility state. If a dispatched job triggers an OOM, the failure detector routes it to recovery (\stepnum{9}), which strengthens the memory requirement and requeues the job for prioritized retry.

Together, these steps form a runtime control loop that decides when a GPU can receive additional work, where that work should run, and how failed placements should be retried.

\begin{algorithm}[t]
\caption{\sysname{} online scheduling with observation holds and runtime-pressure filtering}
\label{alg:aegis-scheduler}
\begin{algorithmic}[1]
\Require Primary queue $Q$, recovery queue $Q_{\mathrm{rec}}$, GPUs $\mathcal{G}$

\While{\textbf{true}}
\State Wait until $Q_{\mathrm{rec}} \neq \emptyset$ or $Q \neq \emptyset$
\State $j \gets \Call{SelectNextJob}{Q_{\mathrm{rec}}, Q}$
\State $k_j \gets \Call{RequiredGPUCount}{j}$
\State $r_j \gets \Call{MemoryRequirement}{j}$
\Statex \Comment{$r_j$ is used only by policies that apply memory feasibility}
\State $\mathcal{S} \gets \Call{GetPerGPUState}{\mathcal{G}}$
\Statex \Comment{$\mathcal{S}$ contains observation-hold status, available memory, runtime-pressure eligibility, and pressure metadata}

\State $\mathcal{C}_{\mathrm{ready}} \gets
    \Call{ExcludeObservationHold}{\mathcal{G}, \mathcal{S}}$

\State $\mathcal{C} \gets
    \Call{BuildCandidateSet}{j, \mathcal{C}_{\mathrm{ready}}, \mathcal{S}, r_j}$
\Statex \Comment{Uses the memory and runtime-pressure information required by the configured policy}

\If{$|\mathcal{C}| \geq k_j$}
    \State $\mathcal{A} \gets
        \Call{PlacementPolicy}{j, \mathcal{C}, \mathcal{S}}$
    \State \Call{Launch}{$j, \mathcal{A}$}
    \State \Call{StartObservationHold}{$j, \mathcal{A}$}
\Else
    \State Keep $j$ pending until sufficient candidate GPUs become available
\EndIf

\If{\Call{FailureDetected}{$j$}}
    \State \Call{RecoverAndRequeue}{$j, Q_{\mathrm{rec}}$}
\EndIf
\EndWhile
\end{algorithmic}
\end{algorithm}

\subsection{Post-Placement Observation and Runtime-Pressure Eligibility}
\label{sec:design:eligibility}

\sysname{} separates a mandatory post-placement observation hold from runtime-pressure eligibility. After launching a job on GPU $g$, the scheduler immediately places $g$ on observation hold and excludes it from further collocation, regardless of the selected placement policy. Existing jobs continue running, but no additional job may be placed on $g$ until the newly launched workload has been observed using the GPU and sufficient post-activity telemetry has been collected. This prevents subsequent placement decisions from using stale memory and pressure information measured before the new workload became active. The observation hold is mandatory and independent of the placement policy, ensuring that every subsequent collocation decision uses post-activity GPU state.

\sysname{} detects the newly launched workload's first observed GPU activity through low-overhead process-level monitoring with \texttt{nvidia-smi}, rather than exact kernel-launch instrumentation. Once activity is observed, \sysname{} samples telemetry once per second into a 30\,s activity-anchored window. The GPU remains on observation hold until this initial window is complete.

When the initial window completes, \sysname{} records its latest available-memory sample, computes the pressure indicators described in \Cref{sec:design:pressure_indicators}, releases the observation hold, and applies the saturation rule. Thereafter, the 30\,s window advances once per second, refreshing available memory, pressure indicators, and runtime-pressure eligibility as samples enter and expire. A GPU remains runtime-pressure ineligible while the saturation rule holds and becomes eligible again when it no longer does.

Anchoring observation at first GPU activity prevents initialization, data loading and preprocessing, CUDA context creation, and early memory allocation from being mistaken for sustained execution headroom. A fixed launch-relative delay is brittle because time to meaningful GPU activity varies across workloads. \Cref{sec:eval:monitoring_window} shows that the selected 30\,s window remains close to a 200\,s reference while enabling responsive scheduler updates.

\subsection{Memory Feasibility}
\label{sec:design:memory}

For memory-aware policies, memory feasibility is the first candidate-filtering stage after observation-hold exclusion. A GPU that does not satisfy the job's memory requirement is removed before runtime-pressure filtering and policy ranking. Memory feasibility determines whether a placement can fit, while runtime-pressure eligibility determines whether an otherwise feasible GPU should receive additional work.

\sysname{} supports three memory-input modes. Profile-backed deployments can use measured solo peak memory, while estimator-backed deployments can use workload-specific GPU memory estimates from analytical models~\cite{horus,gao2020estimating,kim2024llmem}, CPU-side analysis~\cite{faketensor2023,deepspeed_memory_requirements,10.1145/3721462.3770773}, or learned predictors~\cite{yousefzadehGPUMemoryEstimation,gao2023runtime}, which offer different tradeoffs~\cite{yousefzadehGPUMemoryEstimation}. In estimator-free mode, each candidate GPU must satisfy a deployment-configured minimum-free-memory floor before the first placement attempt. All three modes use the same memory-feasibility check; only the source of the memory requirement differs.

A placement may still trigger an OOM because of estimation error, allocator behavior, or runtime peaks not captured by the selected memory requirement. Recovery then strengthens the requirement for subsequent attempts. In all modes, \sysname{} adds a configurable safety margin before checking memory feasibility.

\subsection{Runtime-Pressure Indicators and Saturation Rule}
\label{sec:design:pressure_indicators}

After the mandatory post-placement observation hold completes, \sysname{} uses pressure indicators to determine whether the sufficiently observed GPU is runtime-pressure eligible for further collocation or should remain excluded as saturated. These indicators do not replace memory feasibility or the observation hold: memory feasibility enforces capacity constraints, the hold prevents decisions based on stale post-placement state, and the saturation rule governs subsequent pressure-based collocation.

Our default implementation uses three DCGM profiling metrics~\cite{dcgm,yousefzadeh2023profiling}: SMACT (SM active), SMOCC (SM occupancy), and DRAMA (DRAM active). SMACT measures the fraction of cycles in which an SM has at least one active warp, SMOCC measures resident-warp occupancy relative to the theoretical maximum, and DRAMA measures the fraction of cycles in which the DRAM interface is active~\cite{dcgm-field-ids}.

\sysname{} samples the three metrics over the activity-anchored 30\,s window described in \Cref{sec:design:eligibility}, updating the pressure indicators as new samples arrive and older samples expire.

Let $\mathcal{M}$ denote the metrics SMACT, SMOCC, and DRAMA.
For each metric $m \in \mathcal{M}$, \sysname{} computes a window-level pressure indicator $R_m$. Each indicator combines complementary statistics: the mean captures sustained pressure, the median reduces sensitivity to short spikes, the p95 captures tail pressure, and exponentially weighted moving average (EWMA) emphasizes recent behavior:
\[
R_m=\frac{1}{4}\Big(
\overline{m}+\widetilde{m}+p_{95}(m)+\mathrm{EWMA}_{\alpha}(m)
\Big),
\]
where $\overline{m}$ and $\widetilde{m}$ denote the window mean and median, $W$ is the number of samples in the window, and $\alpha=2/(W+1)$.

After the initial activity-anchored window is complete, a GPU is considered saturated when
\[
R_{\mathrm{SMACT}} \ge \tau_s
\;\land\;
\left(
R_{\mathrm{SMOCC}} \ge \tau_o
\;\lor\;
R_{\mathrm{DRAMA}} \ge \tau_d
\right)
\]
High SMACT alone does not imply saturation: when both SMOCC and DRAMA remain low, the GPU may still have room for additional parallelism. \sysname{} therefore filters a GPU only when high execution activity is accompanied by high occupancy or high DRAM activity. A saturated GPU is runtime-pressure ineligible for policies that apply pressure filtering; when the rule no longer holds, it becomes runtime-pressure eligible again. The thresholds are configuration parameters rather than hardware constants; \Cref{sec:eval:threshold_sensitivity} evaluates their sensitivity and selects the operating point used in our experiments.

\paragraph{Telemetry fallback.}
When the SMACT, SMOCC, and DRAMA metrics are unavailable on the target GPU or through the telemetry stack, \sysname{} falls back to coarse GPU utilization samples from \texttt{nvidia-smi}. It aggregates these samples over the same activity-anchored window to produce a utilization indicator and marks the GPU saturated when that indicator exceeds $\tau_u$. Although this fallback cannot distinguish occupancy from DRAM pressure, it preserves memory feasibility, the mandatory activity-anchored observation hold, and runtime-pressure eligibility derived from the utilization indicator.

\subsection{Placement Policies}
\label{sec:design:policies}
Placement policies operate after GPUs under the mandatory post-placement observation hold are excluded and use the scheduler-maintained per-GPU state according to their design. Policies may additionally apply memory feasibility and runtime-pressure filtering before ranking the remaining candidates.

\textbf{Exclusive} assigns a job only to idle GPUs and disables collocation. This policy represents the conventional one-job-per-GPU execution model and serves as the isolation-oriented baseline.

\textbf{Round-Robin (RR).} It preserves the mandatory observation hold but does not use memory or runtime-pressure information for GPU selection. It assigns each job to the next GPU(s) in circular order among those not under observation hold.

\textbf{Max-Free-Memory (MFM).} It ranks the memory-feasible GPUs that are not under observation hold by their latest available-memory value after runtime-pressure filtering, when enabled, selecting the GPU with the largest available memory.

\textbf{Min-Utilization (MU).} It ranks the memory-feasible GPUs that are not under observation hold according to their latest SMACT-based runtime-pressure indicator after runtime-pressure filtering, when enabled, preferring the GPU with the lowest recent compute pressure.

The placement layer is modular and can support additional policies for locality, fairness, estimated slowdown, or application-specific objectives while preserving mandatory post-placement observation and OOM-aware recovery and selectively applying memory feasibility and runtime-pressure filtering according to policy design.

\subsection{Failure Detection and Recovery}
\label{sec:design:recovery}

Imperfect memory information can still admit placements that prove memory-infeasible at runtime. \sysname{} therefore treats post-dispatch OOM failures as placement-level feedback: the job is requeued with a strengthened memory requirement and falls back to exclusive execution if needed. Such failures can result from underestimated demand, allocator behavior, or runtime allocations that expose a higher peak than expected.

The post-placement observation hold does not prevent the newly launched job's first OOM. Instead, it prevents additional workloads from being colocated using stale pre-placement memory and pressure state while that job is still starting. This limits cascading premature placements and makes the failed placement easier to attribute and recover. This separation allows \sysname{} to treat OOMs as memory-feasibility mistakes rather than errors in the saturation decision.
 
The failure detector classifies terminated jobs using their execution status and runtime logs already captured by the launcher; it does not instrument the GPU execution path. OOM failures reported in the training framework's runtime logs are routed to the recovery manager, while non-OOM failures are recorded but not retried because they may reflect application bugs, data errors, missing dependencies, or other causes unrelated to GPU-memory feasibility.

OOM-affected jobs enter a dedicated recovery queue. While this queue is nonempty, the job selector dequeues from it rather than from the primary queue, ensuring that recovery attempts are handled before newly submitted jobs.

Recovery proceeds through a configurable ladder. The ladder may use fixed memory levels or capacity-aware levels derived from the target GPU’s device memory. In both cases, \sysname{} selects the smallest recovery level that is strictly more conservative than the effective failed threshold. If such a level exists, the job is requeued with the strengthened memory requirement. If no higher level exists, \sysname{} marks the job for full-GPU fallback, causing the scheduler to wait until the job can run exclusively. If a job fails again after full-GPU fallback, recovery stops and the failure is recorded as terminal.

Algorithm~\ref{alg:recovery} summarizes the OOM-aware recovery procedure.

\begin{algorithm}[t]
\caption{\sysname{} OOM-aware recovery}
\label{alg:recovery}
\begin{algorithmic}[1]
\Require Failed job $j$, recovery queue $Q_{\mathrm{rec}}$

\State $r \gets \Call{ClassifyFailure}{j}$

\If{$r \neq \textsc{OOM}$}
\State \Call{RecordNonRecoverableFailure}{$j, r$}
\State \Return
\EndIf

\If{\Call{FailedAfterFullGPUFallback}{$j$}}
\State \Call{RecordTerminalRecoveryFailure}{$j$}
\State \Return
\EndIf

\State $T_{\mathrm{base}} \gets \Call{CurrentMemoryRequirement}{j}$
\State $T_{\mathrm{fail}} \gets \max(T_{\mathrm{base}}, \Call{FailedHostFreeMemory}{j})$
\State $T_{\mathrm{next}} \gets \Call{NextRecoveryLevelAbove}{T_{\mathrm{fail}}}$

\If{$T_{\mathrm{next}}$ exists}
\State \Call{SetRecoveryMemoryRequirement}{$j, T_{\mathrm{next}}$}
\Else
\State \Call{MarkForFullGPUFallback}{$j$}
\EndIf

\State \Call{Enqueue}{$Q_{\mathrm{rec}}, j$}
\end{algorithmic}
\end{algorithm}

\subsection{Default Configuration}
\label{sec:design:default}

Unless configured otherwise, \sysname{} uses MFM placement, mandatory activity-anchored post-placement observation, saturation-based runtime-pressure filtering, and OOM-aware recovery. The default configuration is estimator-free: before a job's first placement attempt, \sysname{} does not require workload-specific profiling or a GPU-memory estimate and instead applies a fixed 5~GB minimum-free-memory precondition.

MFM preserves memory headroom by selecting the remaining candidate with the most available memory. If a placement triggers an OOM, recovery strengthens the memory requirement for subsequent attempts as described in \Cref{sec:design:recovery}. The recovery ladder uses capacity-relative levels at 25\%, 50\%, and 75\% of GPU memory, with intermediate levels limiting each increase to 8\,GiB. The runtime-pressure eligibility thresholds are selected empirically in \Cref{sec:eval:threshold_sensitivity}.

In our experiments, \sysname{} uses MPS-enabled execution for GPU collocation. The evaluated configurations do not dynamically create, resize, or merge MIG partitions.

\section{Evaluation}
\label{sec:eval}

We evaluate whether \sysname{} improves multi-GPU server efficiency through runtime-guided controlled collocation across production-derived traces with heterogeneous deep learning workloads. We compare it with exclusive execution and representative profile- and prediction-guided baselines, then isolate the effects of memory input, placement policy, runtime-pressure filtering, limited telemetry, monitoring-window length, threshold choice, and arrival order. Unless otherwise stated, \sysname{} denotes the default estimator-free configuration in \Cref{sec:design:default}; alternative memory-input and placement variants are evaluated separately.

We ask whether \sysname{}: (1) improves end-to-end performance, GPU utilization, and GPU energy-to-solution relative to baselines; (2) trades reduced queueing for increased execution time and per-task slowdown; (3) remains effective across memory inputs, placement policies, pressure filtering, and limited telemetry; and (4) is robust to monitoring-window length, threshold choice, and arrival order.










\subsection{Experimental Setup}
\label{sec:eval:setup}

\paragraph{Testbeds.}
Our main evaluation runs on an NVIDIA DGX Station with three NVIDIA A100-SXM4-40GB GPUs available for training. The server has one AMD EPYC 7742 processor with 64 cores and 128 hardware threads, 512\,GB of system memory, NVIDIA driver 580.159.03, and CUDA 13.0. We use this platform to study \sysname{} under controlled multi-GPU queueing, collocation, interference, and recovery behavior.

To evaluate \sysname{} beyond the primary A100 testbed, we also use a heterogeneous GPU server with limited telemetry support. The server contains four NVIDIA RTX 2080 Ti GPUs and four NVIDIA GTX 1080 Ti GPUs, each with 11\,GB of device memory, two Intel Xeon Silver 4114 CPUs, 40 hardware threads, 579\,GB of system memory, NVIDIA driver 550.144.03, and CUDA 12.4. Unlike the A100 testbed, this platform does not expose the full DCGM pressure vector uniformly across devices. We therefore evaluate the limited-telemetry fallback described in \Cref{sec:design:pressure_indicators}. It preserves memory-feasibility checks and the mandatory activity-anchored observation hold, but replaces pressure indicators derived from the DCGM metrics SMACT, SMOCC, and DRAMA with a utilization indicator derived from the coarse-grained GPU-utilization metric exposed by \texttt{nvidia-smi}. Because each GPU has only 11\,GB of memory, we restrict this experiment to workloads that are individually feasible on a single device. This setting is not intended to stress OOM recovery; instead, it tests whether activity-anchored observation and runtime-pressure filtering remain useful with coarser telemetry on a heterogeneous server with more GPUs and different GPU generations than the primary A100 testbed.

\paragraph{Workloads.}

Our workload suite spans image classification, segmentation and detection, language modeling, recommendation, and LLM-style training, including single- and two-GPU jobs with diverse batch sizes, model sizes, execution times, memory footprints, and runtime-pressure profiles. We map these workloads onto production-derived arrival traces to evaluate scheduling under realistic job-arrival patterns.

\Cref{tab:workload-summary} summarizes the concrete workload groups, datasets, batch-size ranges, GPU-count requirements, and profiled memory ranges used in the evaluation. Peak GPU memory ranges from 0.6\,GB to 32.9\,GB, creating placements that differ in memory footprint, GPU count, and runtime pressure. This mix follows prior GPU sharing and scheduling evaluations, which commonly combine vision, Transformer/NLP, recommendation, and other non-CNN workloads to expose differences in memory pressure, compute utilization, and interference behavior~\cite{lucid_asplos}. We use the suite as a representative set of training jobs for evaluating memory feasibility, eligibility filtering, and placement decisions under heterogeneous resource demand.

\begin{table}[t]
\centering
\caption{Summary of workloads used in the evaluation. Mem. reports the observed/profiled peak GPU memory range in GB across evaluated specifications (BS: batch size).}
\label{tab:workload-summary}
\small
\setlength{\tabcolsep}{2.0pt}
\renewcommand{\arraystretch}{1.0}
\begin{tabularx}{\columnwidth}{@{}cllccc@{}}
\toprule
Group & Model & Dataset & BS & GPUs & Mem. (GB) \\
\midrule
$\ast$ & ResNet 18/34/50      & ImageNet/CIFAR-100 & 32--128 & 1 & 0.6--21.2 \\
$\ast$ & MobileNet    & ImageNet/CIFAR-100 & 32--128 & 1 & 2.4--3.3 \\
$\ast$ & EfficientNet & ImageNet/CIFAR-100 & 32--128 & 1 & 2.6--4.5 \\
$\ast$ & VGG16        & ImageNet           & 32--128 & 1 & 6.8--21.2 \\
$\ast$ & Xception     & ImageNet           & 32--128 & 1 & 6.7--20.7 \\
$\ast$ & Inception    & ImageNet           & 32--128 & 1 & 5.5--17.0 \\
$\diamond$ & U-Net     & VOC                & 8       & 1 & 5.5 \\
$\diamond$ & Mask R-CNN & COCO               & 8       & 1 & 29.2 \\
$\blacktriangle$ & BERT base/large  & WikiText & 2--32 & 1--2 & 8.7--24.0 \\
$\blacktriangle$ & XLNet base/large & WikiText & 2--32 & 1--2 & 10.2--25.6 \\
$\blacktriangle$ & GPT-2 large/XL   & WikiText & 2--32 & 1--2 & 14.5--32.9 \\
$\star$ & DLRM         & Criteo             & 32768   & 1 & 1.3 \\
$\bullet$ & LLaMA3-width & WikiText         & 1       & 1 & 28.0 \\
\bottomrule
\end{tabularx}
\vspace{0.3em}
\footnotesize
$\ast$ image classification;
$\diamond$ segmentation/detection;
$\blacktriangle$ language/Transformer;
$\star$ recommendation;
$\bullet$ LLM-style.
ImageNet~\cite{imagenet}, CIFAR-100~\cite{cifar}, VOC~\cite{pascal-voc}, COCO~\cite{coco}, WikiText~\cite{wikitext}, and Criteo~\cite{criteo}.
\end{table}

\paragraph{Traces.}
We evaluate three 60-job workload traces derived from the Philly trace~\cite{philly_traces,AnalysisOfMultiTenantGPUClusters} and the Saturn and Venus traces from Helios~\cite{hu2021characterization,heliosdata}. We use our workload-to-trace mapping tool\footnote{The workload-to-trace mapping tool is available at \url{https://github.com/ehsanyousefzadehasl/dl-trace-mapper}.} to map production-trace arrival times to concrete deep learning workload specifications. The resulting traces differ in workload mix and arrival order, allowing us to evaluate \sysname{} under different queueing and collocation conditions.

For the limited-telemetry server, we use the same mapping procedure with an 11\,GB-feasible workload catalog.

\paragraph{Baselines.}
We compare \sysname{} with exclusive execution and representative collocation baselines. Exclusive execution assigns jobs only to idle GPUs and disables collocation. Horus~\cite{horus} is a prediction-guided baseline that uses an analytical model to estimate GPU-memory demand and a learned model to predict workload-level GPU utilization as an interference proxy for placement. In our evaluation, Horus receives the measured utilization associated with each workload specification, representing an optimistic setting for its utilization input. Lucid~\cite{lucid_asplos} receives workload memory requirements, features extracted from 200\,s profiles, and the resulting classifier-assigned compatibility labels before trace execution begins.

For both baselines, we exclude setup from trace makespan and JCT, so reported end-to-end metrics cover only scheduling and execution after the trace begins. The corresponding inputs, profiling artifacts, labels, and configurations are included with the evaluation artifacts. In contrast, \sysname{} uses memory-feasibility checks, mandatory activity-anchored observation, runtime-pressure filtering, and OOM-aware recovery during scheduling without requiring offline pairwise compatibility profiling.

\paragraph{\sysname{} setup.}
The default \sysname{}, denoted \sysname{}-EstimatorFree in the evaluation for clarity, configuration is estimator-free: it uses the MFM placement policy, a fixed 5\,GB minimum-free-memory precondition before the first placement attempt, the mandatory activity-anchored post-placement observation hold, saturation-based runtime-pressure filtering with the thresholds selected in \Cref{sec:eval:threshold_sensitivity}, and OOM-aware recovery.

We also evaluate \sysname{} variants that change the memory input used for feasibility checks. \sysname{}+PeakMem uses profiled peak GPU memory, while \sysname{}+AnalyticalMem uses a lightweight analytical memory estimator inspired by Horus~\cite{horus}; additional estimator variants are analyzed in \Cref{sec:eval:memory_estimation}. To isolate scheduler mechanisms, we further vary the placement policy and include no-filter variants that disable saturation-based pressure filtering while preserving memory feasibility and the mandatory observation hold. 

\paragraph{Metrics.}
We report server-level and job-level metrics. Server-level metrics include trace makespan, GPU utilization, and GPU energy-to-solution, defined as the sum of energy consumed by all training GPUs from trace start until all jobs complete. Job-level metrics include job completion time (JCT), queueing time, execution time, slowdown relative to solo execution, OOM failures, and recovered attempts. Collocation should primarily reduce makespan and queueing while increasing GPU utilization, although interference may lengthen some jobs' execution time. Queueing time includes recovery wait, and execution time includes failed-attempt runtime. We report within-trace p95 job percentiles rather than p99 because each trace contains 60 jobs; p95 captures tail behavior without being effectively determined by one or two jobs.

\paragraph{Monitoring and thresholds.}
\sysname{} samples GPU telemetry once per second. After each placement, the affected GPU enters mandatory observation hold until the newly launched workload's first observed GPU activity and a complete 30\,s post-activity window have been collected. The completed window refreshes the latest available-memory value, computes the first pressure indicators, and applies the saturation rule to establish runtime-pressure eligibility. Thereafter, the window advances once per second, updating available memory and pressure indicators and refreshing eligibility as new metric samples arrive and older samples expire. The default DCGM metrics are SMACT, SMOCC, and DRAMA; their derived pressure indicators are compared against thresholds $\tau_s=0.65$, $\tau_o=0.35$, and $\tau_d=0.50$, respectively. \Cref{sec:eval:threshold_sensitivity} studies threshold sensitivity and selects this operating point.
On the heterogeneous GTX/RTX testbed, where the SMACT, SMOCC, and DRAMA metrics are unavailable through our telemetry stack, \sysname{} instead derives a utilization indicator from the coarse-grained GPU-utilization metric exposed by \texttt{nvidia-smi}. In this fallback mode, a GPU is marked saturated when the utilization indicator exceeds $\tau_u=70\%$.

\subsection{End-to-End Performance Against Baselines}

We first evaluate whether \sysname{} improves end-to-end performance across workload traces relative to exclusive execution and the representative collocation baselines described in \Cref{sec:eval:setup}. Unless otherwise stated, \sysname{}-EstimatorFree denotes the default configuration defined there. The normalized comparisons also include profiled-peak and analytical memory-input variants, which differ only in memory source; \Cref{sec:eval:memory_estimation} analyzes them in detail. Relative to Exclusive, Horus, and Lucid, \sysname{}-EstimatorFree achieves the lowest makespan, mean and p95 JCT, and mean and p95 queueing time on all three traces.

\begin{figure*}[t]
\centering
\begin{subfigure}[t]{0.485\linewidth}
\centering
\includegraphics[width=\linewidth]{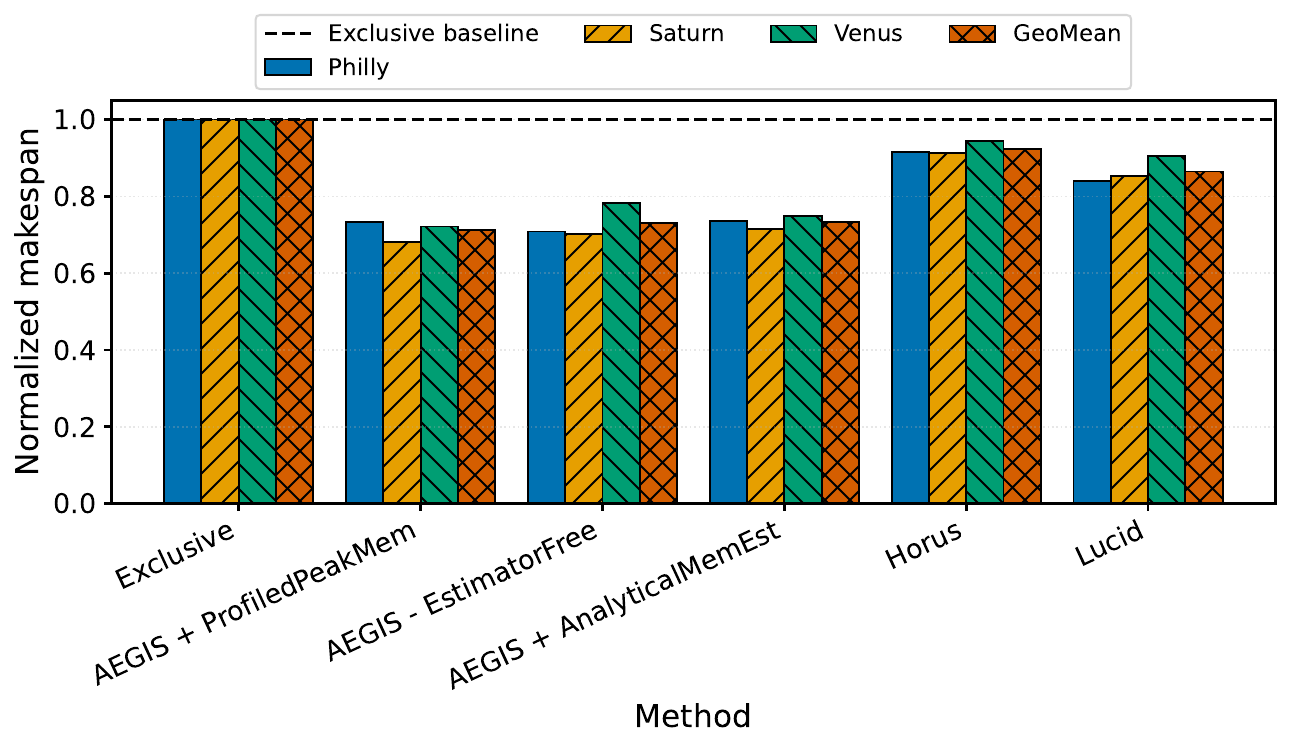}
\caption{Normalized makespan.}
\label{fig:eval-normalized-makespan}
\end{subfigure}
\hfill
\begin{subfigure}[t]{0.485\linewidth}
\centering
\includegraphics[width=\linewidth]{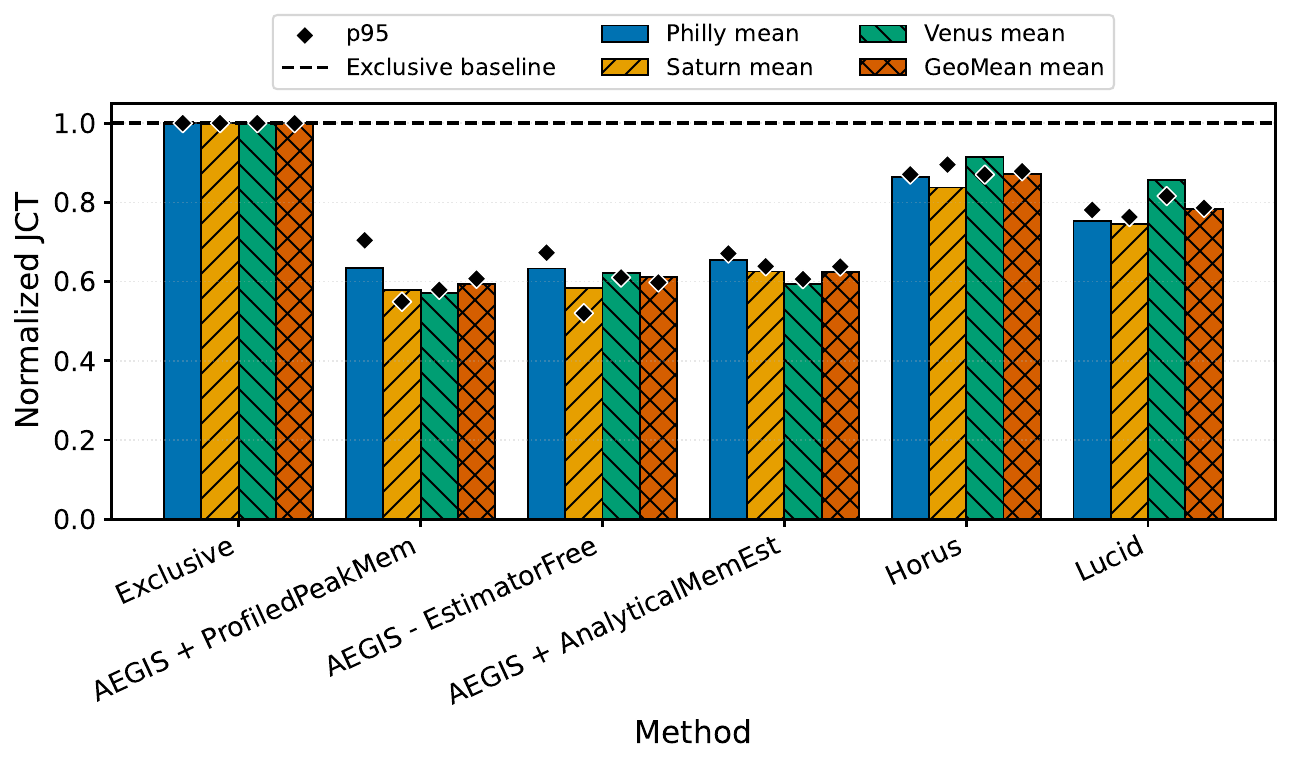}
\caption{Normalized mean JCT.}
\label{fig:eval-normalized-jct}
\end{subfigure}
\caption{
End-to-end trace performance. Values are normalized to Exclusive on the corresponding trace; lower is better. Bars show per-trace values and GeoMean. In the JCT panel, diamonds show p95 JCT normalized to Exclusive p95 JCT on the same trace. Results include failed-attempt runtime and recovery wait.
}
\label{fig:eval-end-to-end}
\end{figure*}

Across traces, \sysname{}-EstimatorFree reduces GeoMean makespan by 27.0\% relative to Exclusive, 21.0\% relative to Horus, and 15.6\% relative to Lucid. It also reduces GeoMean JCT by 38.7\% relative to Exclusive, 29.7\% relative to Horus, and 21.8\% relative to Lucid. \Cref{tab:eval-raw-end-to-end} reports the corresponding absolute values in hours, while \Cref{fig:eval-end-to-end} shows the normalized results.

\begin{table}[t]
\centering
\caption{
Absolute end-to-end performance across traces. Values are hours; p95 values are within-trace job percentiles. Lower is better for all metrics; bold marks the best value in each row.
}
\label{tab:eval-raw-end-to-end}
\small
\setlength{\tabcolsep}{3.4pt}
\renewcommand{\arraystretch}{0.92}
\begin{tabular}{llrrr|r}
\toprule
\textbf{Metric (h)} & \textbf{\textit{Trace}} & \textbf{Exclusive} & \textbf{Horus} & \textbf{Lucid} & \textbf{\shortstack{\sysname{}\\Est.-Free}} \\
\midrule
\multirow{3}{*}{Makespan}
& \textit{Philly} & 11.84 & 10.84 & 9.94 & \textbf{8.39} \\
& \textit{Saturn} & 13.17 & 12.01 & 11.22 & \textbf{9.24} \\
& \textit{Venus}  & 12.35 & 11.67 & 11.17 & \textbf{9.66} \\
\midrule
\multirow{3}{*}{Mean JCT}
& \textit{Philly} & 5.23 & 4.52 & 3.93 & \textbf{3.31} \\
& \textit{Saturn} & 4.78 & 4.00 & 3.56 & \textbf{2.79} \\
& \textit{Venus}  & 4.55 & 4.16 & 3.90 & \textbf{2.83} \\
\midrule
\multirow{3}{*}{P95 JCT}
& \textit{Philly} & 8.54 & 7.44 & 6.67 & \textbf{5.75} \\
& \textit{Saturn} & 11.37 & 10.17 & 8.67 & \textbf{5.91} \\
& \textit{Venus}  & 6.85 & 5.96 & 5.59 & \textbf{4.18} \\
\midrule
\multirow{3}{*}{Mean Queue}
& \textit{Philly} & 4.70 & 3.98 & 3.34 & \textbf{2.41} \\
& \textit{Saturn} & 4.23 & 3.45 & 2.96 & \textbf{1.86} \\
& \textit{Venus}  & 4.02 & 3.61 & 3.34 & \textbf{2.15} \\
\midrule
\multirow{3}{*}{P95 Queue}
& \textit{Philly} & 8.09 & 6.93 & 6.16 & \textbf{4.41} \\
& \textit{Saturn} & 10.22 & 9.03 & 8.23 & \textbf{5.56} \\
& \textit{Venus}  & 6.62 & 5.70 & 5.27 & \textbf{3.64} \\
\midrule
\multirow{3}{*}{Mean Job Exec.}
& \textit{Philly} & \textbf{0.53} & 0.54 & 0.59 & 0.90 \\
& \textit{Saturn} & \textbf{0.55} & 0.56 & 0.59 & 0.93 \\
& \textit{Venus}  & \textbf{0.54} & 0.55 & 0.57 & 0.69 \\
\midrule
\multirow{3}{*}{P95 Job Exec.}
& \textit{Philly} & \textbf{0.98} & 1.13 & 1.16 & 2.07 \\
& \textit{Saturn} & \textbf{0.99} & 1.02 & 1.15 & 2.54 \\
& \textit{Venus}  & \textbf{1.00} & 1.14 & 1.12 & 1.42 \\
\bottomrule
\end{tabular}
\end{table}

\begin{figure*}[t]
  \centering
  \includegraphics[width=\textwidth]{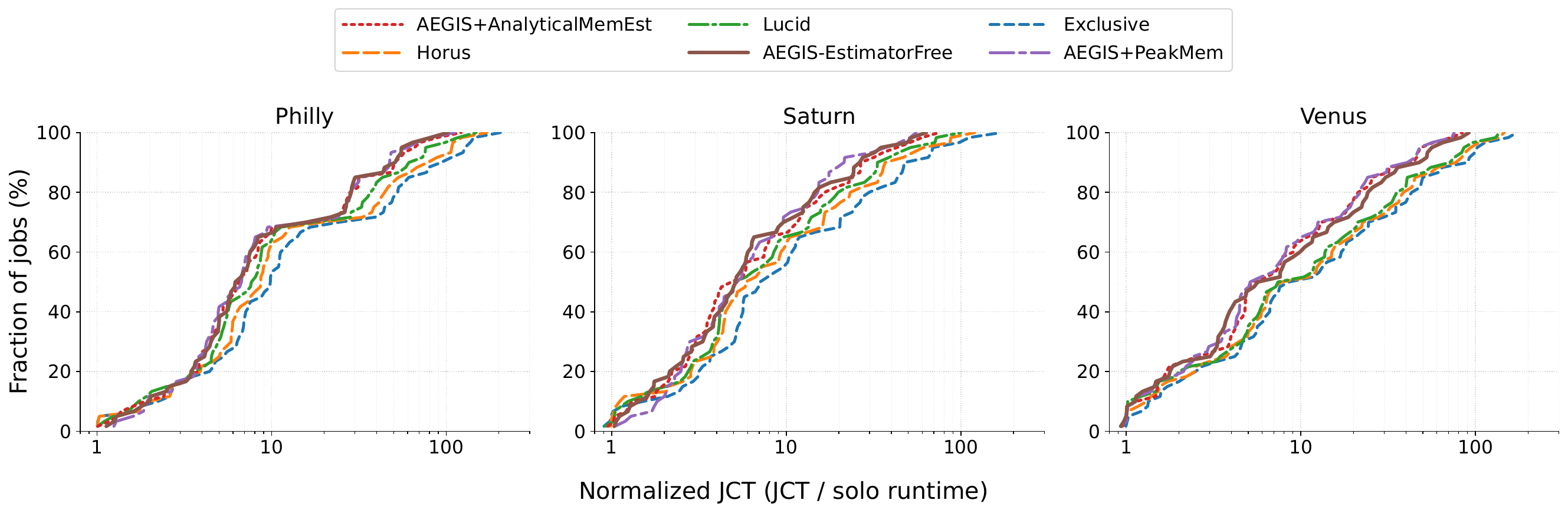}
  \caption{
    Per-job normalized JCT distributions across traces. Each job's JCT is divided by its isolated solo runtime; curves farther left are better. The x-axes are logarithmic.
  }
  \label{fig:eval-jct-cdf}
\end{figure*}

\label{sec:eval:end_to_end}

These gains come primarily from queueing reduction: mean queueing time falls to $0.494\times$ Exclusive and p95 queueing time to $0.546\times$. All reported results include failed-attempt runtime and recovery wait; \sysname{}-EstimatorFree recovered 3, 3, and 6 jobs in Philly, Saturn, and Venus, respectively.

The cost is longer execution time: as \Cref{tab:eval-raw-end-to-end} shows, \sysname{}-EstimatorFree increases mean job execution time on all traces. Overall, \sysname{} improves makespan and response time by starting jobs earlier on memory-feasible, runtime-pressure eligible GPUs, although some jobs experience slowdown, particularly in the tail. \Cref{sec:eval:jct_breakdown} analyzes this trade-off further.

\Cref{fig:eval-jct-cdf} complements the aggregate results with per-job normalized-JCT distributions. Compared with Exclusive and the collocation baselines, \sysname{}-EstimatorFree shifts the distribution left in most cases. Estimator-backed variants follow similar trends, with differences caused by changes in the memory-feasible candidate set. The remaining tail reflects queueing delay, interference-induced execution-time increases, and trace-dependent differences in concurrent execution.

Resource and energy telemetry attribute the end-to-end gains to higher effective GPU use: across the three traces, \sysname{}-EstimatorFree reduces GPU energy-to-solution by 12.2\% relative to Exclusive while increasing mean GPU utilization by 23.3\%. It also increases mean and p95 active GPU memory usage by 71.0\% and 10.5\%, respectively, indicating that \sysname{} exposes otherwise idle GPU capacity rather than merely shifting work across time.

The higher concurrent GPU activity is also reflected in the metrics: relative to Exclusive, mean SMACT, SMOCC, and DRAMA increase by 39.9\%, 45.1\%, and 39.4\%, respectively. Compared with Horus and Lucid, \sysname{}-EstimatorFree reduces GPU energy-to-solution by 10.5\% and 8.4\% while increasing mean GPU utilization by 14.1\% and 9.0\%, respectively. Relative to Horus, mean SMACT, SMOCC, and DRAMA increase by 35.6\%, 41.1\%, and 24.9\%; relative to Lucid, they increase by 6.3\%, 9.6\%, and 17.3\%, respectively.

\subsection{Queueing--Execution Trade-off}
\label{sec:eval:jct_breakdown}

\begin{figure*}[t]
\centering
\begin{subfigure}[t]{0.485\linewidth}
\centering
\includegraphics[width=\linewidth]{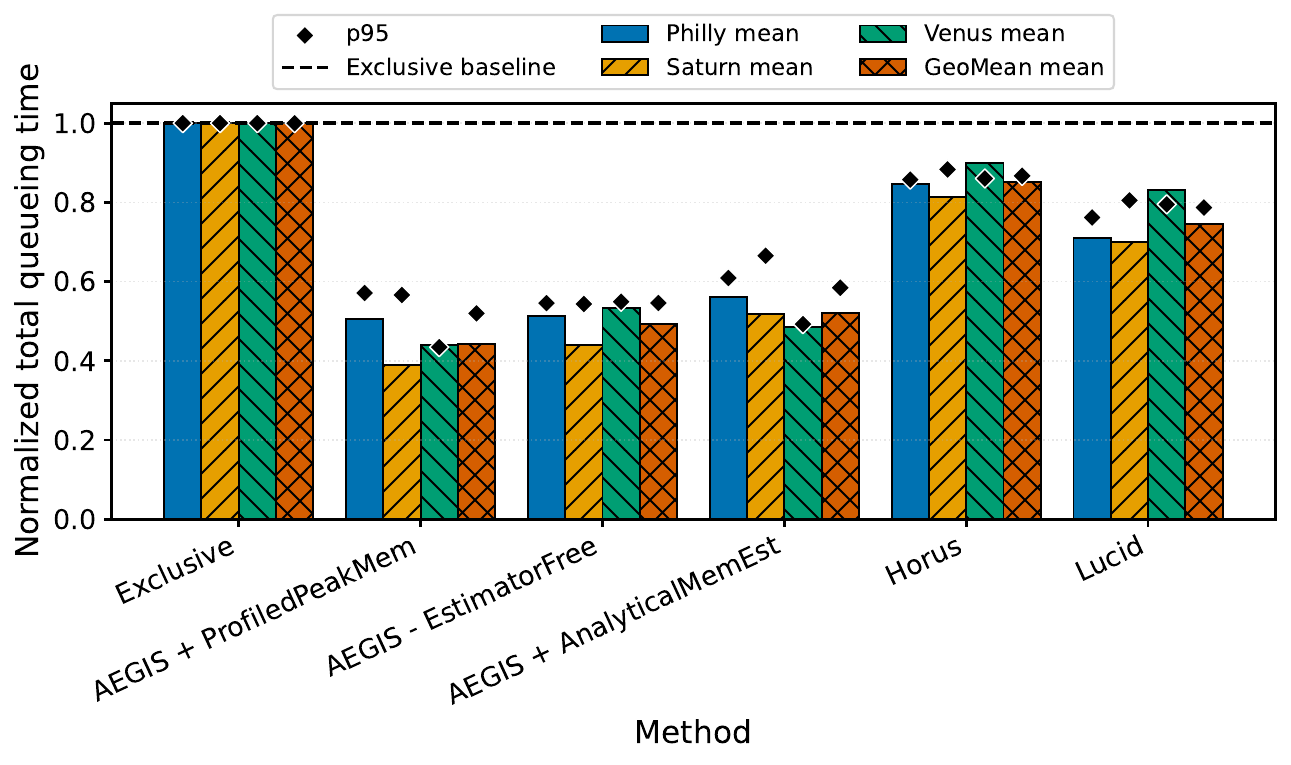}
\caption{Normalized job total queueing time.}
\label{fig:eval-normalized-queue-wait}
\end{subfigure}
\hfill
\begin{subfigure}[t]{0.485\linewidth}
\centering
\includegraphics[width=\linewidth]{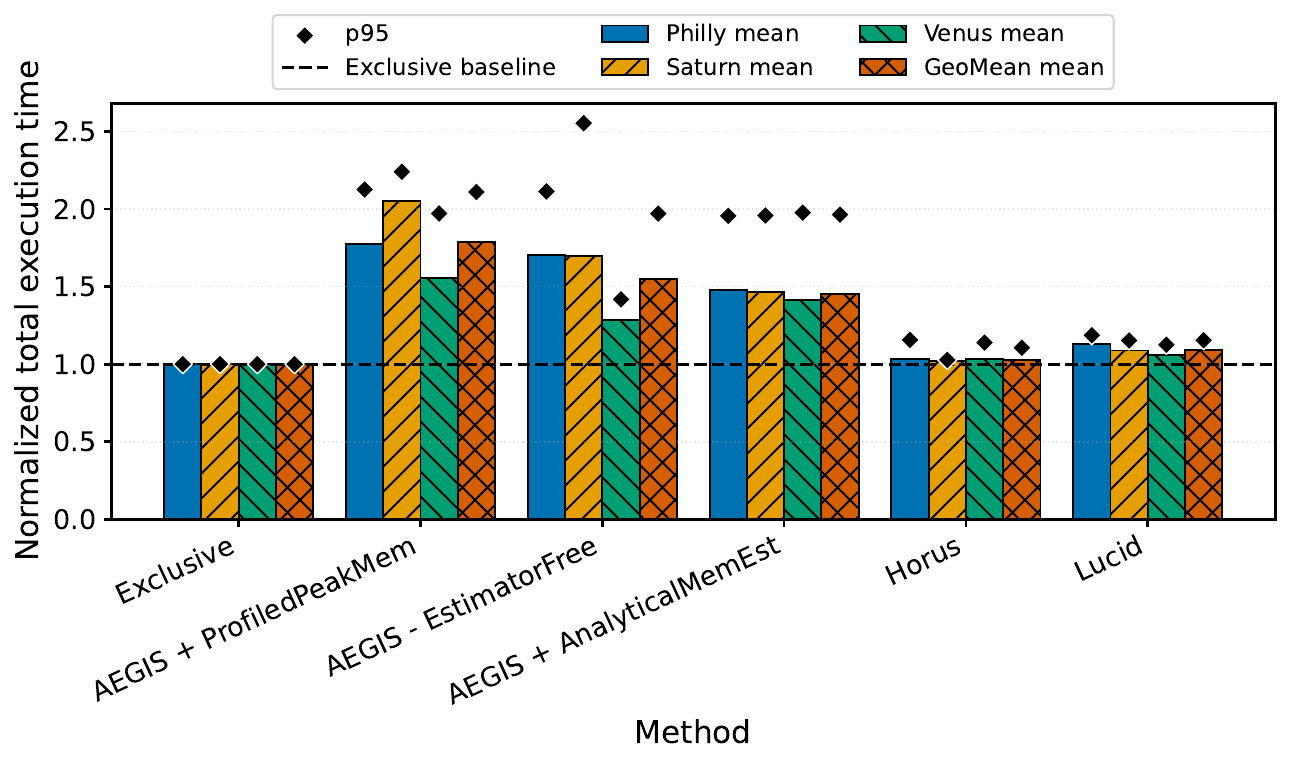}
\caption{Normalized job total execution time.}
\label{fig:eval-normalized-execution-span}
\end{subfigure}
\caption{
Queueing and execution-time trade-off. Values are normalized to Exclusive on the corresponding trace; lower is better. Bars show means. Diamonds show p95 values normalized to the corresponding Exclusive p95 for the same metric and trace.
}
\label{fig:eval-queue-execution-tradeoff}
\end{figure*}

\Cref{tab:eval-raw-end-to-end} reports the absolute queueing and execution-time effects, while \Cref{fig:eval-queue-execution-tradeoff} shows the same decomposition normalized to Exclusive. \sysname{}-EstimatorFree lowers mean and p95 queueing time on every trace, but increases job execution time under collocation.

\Cref{fig:eval-normalized-queue-wait} shows that \sysname{} improves makespan and JCT by placing jobs earlier on already-active GPUs that remain memory-feasible and pressure-eligible, rather than waiting for an idle GPU. This also helps explain the gap to Horus and Lucid: Horus uses a single workload-level utilization prediction as an interference proxy, while Lucid relies on classifier-assigned compatibility labels learned from offline pairwise profiling. These coarser or profile-derived signals can limit execution-time slowdown by avoiding harmful overlaps, but may also reject opportunities that \sysname{} admits using current multidimensional runtime pressure.

\Cref{fig:eval-normalized-execution-span} shows the cost of sharing: some colocated jobs run longer because they contend for SM resources, memory bandwidth, cache capacity. The p95 markers show that this cost is uneven: some jobs overlap briefly or use complementary resources, while others overlap longer or stress the same bottleneck. Overall, \sysname{} reduces mean JCT because the queueing reduction outweighs the increase in execution time, but it does not eliminate tail slowdown.

\subsection{Effect of Memory Input}
\label{sec:eval:memory_estimation}

The memory input determines the memory-feasible candidate set after GPUs under the mandatory activity-anchored post-placement observation hold are excluded and before any enabled runtime-pressure filtering is applied. More accurate or conservative memory inputs can prevent memory-infeasible placements and recovery, but also alter queue evolution by changing the memory-feasible candidate set. We compare them under the same mandatory post-placement observation hold, saturation-based runtime-pressure filter, and MFM placement policy.

\sysname{}-EstimatorFree uses the default estimator-free configuration from \Cref{sec:eval:end_to_end}. PeakMem uses measured solo peak memory, removing memory-estimation error from the feasibility check, but it does not optimize interference, execution-time slowdown, or the schedule produced by placement order and runtime pressure. AnalyticalMem, GPUMemNet~\cite{yousefzadehGPUMemoryEstimation}, and FakeTensor~\cite{faketensor2023} use estimated memory requirements from a lightweight analytical model inspired by Horus~\cite{horus}, a learned memory predictor, and CPU-side FakeTensor analysis, respectively.

\Cref{tab:eval-memory-estimation} reports geometric-mean metrics normalized to Exclusive across the three traces. All configurations complete all jobs and improve makespan and JCT relative to Exclusive, showing that \sysname{} is not tied to a single memory-information source. PeakMem achieves the lowest GeoMean makespan and mean JCT, reducing makespan to $0.712\times$ and mean JCT to $0.594\times$ Exclusive with no failed or recovered attempts. \sysname{}-EstimatorFree remains close, with makespan at $0.730\times$ and mean JCT at $0.613\times$, but triggers 12 failed--recovered attempts. Thus, operating without workload-specific profiles or estimates preserves strong end-to-end performance, but increases the number of failed and recovered attempts. Because training OOMs typically occur during startup or early iterations, these failed placements can often be detected and retried before substantial training progress is lost.

For workloads on which FakeTensor or GPUMemNet does not produce a valid estimate, we conservatively assign a 40\,GB memory requirement~\cite{yousefzadehGPUMemoryEstimation}, effectively preventing collocation on the A100 testbed. This avoids treating missing estimates as small memory demands, but reduces feasible-placement opportunities; a lower fallback would expose more sharing opportunities at the cost of potentially increasing OOM recoveries.

\begin{table}[t]
\centering
\caption{
Effect of memory input. Values are GeoMeans across traces, normalized to Exclusive; lower is better for the time metrics; F/R reports failed/recovered attempts. Bold marks the best value in each metric column; the italic row label marks the default \sysname{} configuration. Queue includes recovery wait; Exec. includes failed-attempt runtime.
}
\label{tab:eval-memory-estimation}
\small
\setlength{\tabcolsep}{2.3pt}
\begin{tabular}{lccccccc|c}
\toprule
& & \multicolumn{2}{c}{JCT} & \multicolumn{2}{c}{Queue} & \multicolumn{2}{c}{Exec.} & \\
\cmidrule(lr){3-4}\cmidrule(lr){5-6}\cmidrule(lr){7-8}
Mem. input & Mkspn & Mean & P95 & Mean & P95 & Mean & P95 & F/R \\
\midrule
PeakMem   & \textbf{0.712} & \textbf{0.594} & 0.607 & \textbf{0.442} & \textbf{0.520} & 1.785 & 2.110 & 0/0 \\
\textit{Est-free}  & 0.730 & 0.613 & \textbf{0.598} & 0.494 & 0.546 & 1.550 & 1.971 & 12/12 \\
AnalyticalMem  & 0.733 & 0.624 & 0.638 & 0.520 & 0.584 & \textbf{1.453} & 1.964 & 3/3 \\
GPUMemNet & 0.737 & 0.629 & 0.636 & 0.510 & 0.573 & 1.575 & 2.057 & 1/1 \\
FakeTensor & 0.794 & 0.648 & 0.679 & 0.543 & 0.641 & 1.483 & \textbf{1.833} & 0/0 \\
\bottomrule
\end{tabular}
\end{table}

The estimator-backed variants show that estimation behavior and coverage both shape schedule quality: FakeTensor produces no recoveries, but has the longest aggregate makespan among them at $0.794\times$ Exclusive. Their incomplete estimator coverage also removes some jobs from collocation, reducing feasible-placement opportunities and reshaping queue evolution.

Memory information shapes both OOM avoidance and scheduling: measured peak memory gives the lowest GeoMean makespan in these traces, but requires prior profiling of representative workload configurations. Practical estimators remain useful because \sysname{} combines memory feasibility with activity-anchored observation, runtime-pressure filtering, and recovery: overestimation and conservative fallbacks reduce collocation opportunities, while underestimation may trigger recovery. Detailed per-workload estimator behavior is included in the evaluation artifacts.

\subsection{Effect of Placement Policy}
\label{sec:eval:placement}

We next evaluate the effect of placement policy (\Cref{sec:design:policies}). All policies preserve the mandatory post-placement observation hold and satisfy each job's GPU-count requirement. MFM and MU additionally use memory feasibility and runtime-pressure information: MFM selects the candidate with the most available memory, while MU selects the one with the lowest SMACT-based pressure. RR ignores both memory and runtime-pressure information and assigns jobs circularly among GPUs not under observation hold.

\Cref{tab:eval-placement} shows that all three policies improve over Exclusive, but the information-aware policies provide a better overall trade-off. MFM achieves the best aggregate makespan ($0.730\times$ Exclusive), lowest p95 JCT ($0.598\times$), lowest mean and p95 execution time ($1.550\times$ and $1.971\times$), and fewest recoveries (12). MU achieves the lowest mean JCT and mean queueing time, but with longer execution time and slightly worse makespan. RR has the highest JCT, queueing time, and recovery count; its 28 failed--recovered attempts reflect circular placement without memory-feasibility or runtime-pressure guidance. We therefore use MFM as the default policy.

\begin{table}[t]
\centering
\caption{
Effect of placement policies. Values are GeoMeans across traces, normalized to Exclusive; lower is better for the time metrics; F/R reports failed/recovered attempts. Bold marks the best value in each metric column; the italic row label marks the default \sysname{} policy. Queue includes recovery wait; Exec. includes failed-attempt runtime.
}
\label{tab:eval-placement}
\small
\setlength{\tabcolsep}{2.8pt}
\begin{tabular}{lccccccc|c}
\toprule
& & \multicolumn{2}{c}{JCT} & \multicolumn{2}{c}{Queue} & \multicolumn{2}{c}{Exec.} & \\
\cmidrule(lr){3-4}\cmidrule(lr){5-6}\cmidrule(lr){7-8}
Policy & Mkspn & Mean & P95 & Mean & P95 & Mean & P95 & F/R \\
\midrule
RR  & 0.776 & 0.635 & 0.636 & 0.501 & 0.573 & 1.686 & 1.993 & 28/28 \\
MU  & 0.755 & \textbf{0.587} & 0.619 & \textbf{0.449} & 0.556 & 1.683 & 2.160 & 13/13 \\
\textit{MFM} & \textbf{0.730} & 0.613 & \textbf{0.598} & 0.494 & \textbf{0.546} & \textbf{1.550} & \textbf{1.971} & 12/12 \\
\bottomrule
\end{tabular}
\end{table}

\subsection{Runtime-Pressure Eligibility Ablation}
\label{sec:eval:pressure_eligibility}

We isolate the effect of runtime-pressure eligibility using no-filter variants that preserve memory feasibility, the mandatory post-placement observation hold, the placement policy, and OOM-aware recovery, but skip saturation-based pressure filtering after the hold is released.

\Cref{tab:runtime-pressure-ablation} shows that disabling pressure filtering worsens the overall performance trade-off. With MFM, makespan changes only slightly, from $0.730\times$ to $0.732\times$ Exclusive, but mean execution time increases from $1.550\times$ to $1.821\times$, p95 execution time from $1.971\times$ to $2.265\times$, and failed--recovered attempts from 12 to 18. Mean queueing time decreases from $0.494\times$ to $0.470\times$, indicating that more permissive placement can start some jobs earlier; however, p95 queueing time increases from $0.546\times$ to $0.570\times$. The higher recovery count likely contributes to this worse tail because affected jobs re-enter the prioritized recovery path. MU shows a similar overall pattern: disabling the filter worsens makespan, mean and p95 JCT, queueing time, execution time, and failed--recovered attempts.

\begin{table}[t]
\centering
\caption{Runtime-pressure eligibility ablation. Values are GeoMeans across traces, normalized to Exclusive; lower is better for the time metrics; F/R reports failed/recovered attempts. Bold marks the best value in each metric column; the italic row label marks the default. Queue includes recovery wait; Exec. includes failed-attempt runtime. NF: saturation-based pressure filtering disabled; the post-placement observation hold remains enabled.}
\label{tab:runtime-pressure-ablation}
\small
\setlength{\tabcolsep}{2.8pt}
\begin{tabular}{lccccccc|c}
\toprule
& & \multicolumn{2}{c}{JCT} & \multicolumn{2}{c}{Queue} & \multicolumn{2}{c}{Exec.} & \\
\cmidrule(lr){3-4}\cmidrule(lr){5-6}\cmidrule(lr){7-8}
Config. & Mkspn & Mean & P95 & Mean & P95 & Mean & P95 & F/R \\
\midrule
\textit{MFM}     & \textbf{0.730} & 0.613 & \textbf{0.598} & 0.494 & \textbf{0.546} & \textbf{1.550} & \textbf{1.971} & 12/12 \\
MFM-NF           & 0.732 & 0.622 & 0.646 & 0.470 & 0.570 & 1.821 & 2.265 & 18/18 \\
MU               & 0.755 & \textbf{0.587} & 0.619 & \textbf{0.449} & 0.556 & 1.683 & 2.160 & 13/13 \\
MU-NF            & 0.770 & 0.616 & 0.626 & 0.480 & 0.580 & 1.740 & 2.250 & 19/19 \\
\bottomrule
\end{tabular}
\end{table}

GPU energy telemetry shows only a marginal difference. MFM-NF reduces GPU energy-to-solution by 12.8\% relative to Exclusive, compared with 12.2\% for thresholded MFM, but this small additional reduction comes with substantially higher execution time and more recovery events. For MU, disabling the filter slightly worsens GPU energy-to-solution.

Overall, saturation-based pressure filtering preserves most of the makespan and GPU energy benefits of collocation while reducing execution-time increases, tail queueing, and recovery events associated with more permissive high-pressure placements.

\subsection{Heterogeneous Limited-Telemetry Testbed}
\label{sec:eval:limited-telemetry}

\Cref{tab:limited-telemetry} evaluates \sysname{} on the heterogeneous RTX/GTX server, where the SMACT, SMOCC, and DRAMA metrics are unavailable through our telemetry stack. \sysname{} therefore derives a utilization indicator from the coarse-grained GPU utilization metric exposed by \texttt{nvidia-smi}. This experiment tests whether the mandatory activity-anchored observation hold and utilization-based saturation rule remain useful with a less expressive runtime signal.

Memory-only sharing serves as an interference-unaware baseline. It nearly eliminates queueing, reducing mean queueing time to $0.116\times$ Exclusive, but increases mean execution time to $1.731\times$. Consequently, mean JCT remains nearly unchanged and p95 JCT worsens to $1.076\times$. Memory feasibility alone is therefore insufficient for beneficial collocation, even when every job is individually feasible on one GPU.

For the \texttt{nvidia-smi}-based fallback, we select $\tau_u=70\%$ from a calibration sweep over $\tau_u \in \{50\%,60\%,70\%,80\%\}$ on the same testbed. The 70\% operating point achieves the best aggregate makespan among the tested thresholds while substantially reducing the execution-time increase observed under memory-only sharing.

The resulting \sysname{}-U70 configuration reduces makespan to $0.884\times$ Exclusive, the best value in the table, while reducing mean JCT to $0.892\times$ and p95 JCT to $0.858\times$. Horus and Lucid achieve slightly lower JCT, but \sysname{}-U70 remains competitive while relying only on the coarse GPU-utilization metric. This suggests that \sysname{} does not depend on the full DCGM pressure vector: the mandatory observation hold still prevents stale-state placements, while the derived utilization indicator provides a useful, though less expressive, signal for the saturation decision.

\begin{table}[t]
\centering
\caption{
Limited-telemetry evaluation on the heterogeneous RTX/GTX server. Values are GeoMeans across traces, normalized to Exclusive; lower is better for the time metrics; F/R reports failed/recovered attempts. Bold marks the best value in each metric column; the italic row label marks the selected \sysname{} fallback. Queue includes recovery wait; Exec. denotes execution time. U70\%: 70\% utilization threshold.
}
\label{tab:limited-telemetry}
\small
\setlength{\tabcolsep}{2.3pt}
\begin{tabular}{lccccccc|c}
\toprule
& & \multicolumn{2}{c}{JCT} & \multicolumn{2}{c}{Queue} & \multicolumn{2}{c}{Exec.} & \\
\cmidrule(lr){3-4}\cmidrule(lr){5-6}\cmidrule(lr){7-8}
Config. & Mkspn & Mean & P95 & Mean & P95 & Mean & P95 & F/R \\
\midrule
AEGIS (Mem-only) & 1.007 & 0.998 & 1.076 & \textbf{0.116} & \textbf{0.155} & 1.731 & 1.955 & 0/0 \\
\textit{AEGIS (U70\%)} & \textbf{0.884} & 0.892 & 0.858 & 0.585 & 0.754 & 1.172 & 1.115 & 0/0 \\
Horus & 0.897 & 0.890 & 0.844 & 0.647 & 0.760 & \textbf{1.112} & 1.044 & 0/0 \\
Lucid & 0.950 & \textbf{0.876} & \textbf{0.832} & 0.510 & 0.631 & 1.195 & \textbf{0.963} & 0/0 \\
\bottomrule
\end{tabular}
\end{table}

The results also show why queueing, JCT, and execution time should be reported separately. Limited-telemetry collocation reduces JCT by shortening queueing time, but the net benefit depends on the resulting execution-time increase, particularly in the tail. This experiment should therefore be interpreted as a deployability and robustness study rather than a replacement for the multidimensional DCGM-based pressure indicators used in the primary evaluation.

\subsection{Monitoring Window and Pressure Indicators}
\label{sec:eval:monitoring_window}

\sysname{} anchors monitoring at first observed GPU activity rather than dispatch time. Across 50 solo workload runs, the median delay from dispatch to first observed GPU activity is 8.2\,s, with a p95 of 20.6\,s and a maximum of 55.5\,s for the LLaMA-family configuration. This variation motivates first-activity anchoring and the mandatory observation hold, since a short, fixed monitoring window starting at dispatch may miss meaningful GPU execution.

After activity begins, \sysname{} must collect enough telemetry to refresh available memory and make the first runtime-pressure classification without extending the mandatory observation hold unnecessarily. We use the 200\,s activity-anchored solo profiles as a stable reference because all evaluated workloads reach sustained execution within this interval. Unlike fixed-duration profiling, however, activity anchoring remains robust when startup delay varies or becomes longer for new workload types.

\begin{figure}[t]
\centering
\includegraphics[width=\linewidth]{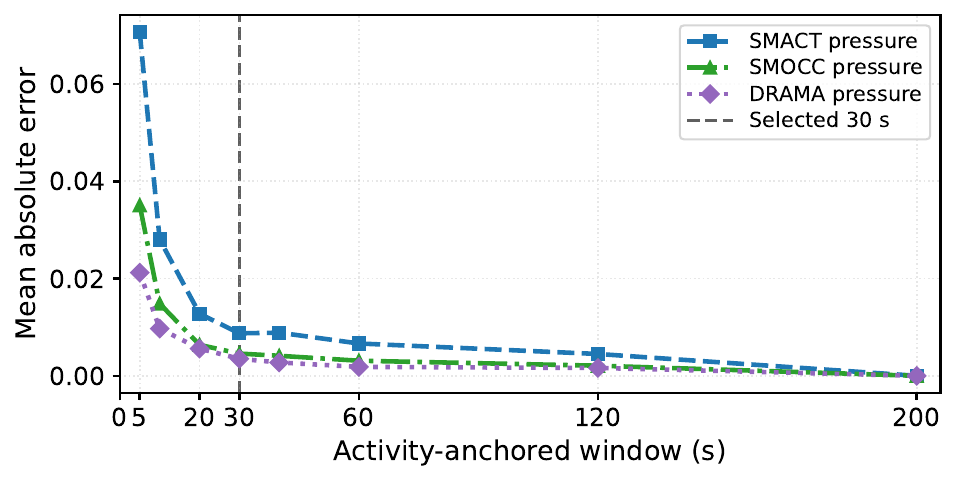}
\caption{
Stability of activity-anchored monitoring windows. Points show weighted mean absolute error of the \sysname{} pressure indicators relative to a 200\,s activity-anchored reference. 
}
\label{fig:window-stability}
\end{figure}

\Cref{fig:window-stability} compares shorter activity-anchored windows with the 200\,s reference. The 30\,s window achieves a weighted mean absolute error of 0.0055 and a weighted mean relative error of 3.31\% across the three pressure indicators. Low error is not the only motivation for this design: each indicator combines mean, median, p95, and EWMA summaries, which capture sustained pressure, robustness to short spikes, tail behavior, and recent activity, respectively. We therefore use 30\,s as the default because it provides a responsive first decision while preserving the complementary pressure characteristics observed over the longer reference.

The activity-anchored window also refreshes available memory. In the solo-profile runs, only 4 of 50 workloads exceed the memory peak observed during the first 30\,s later in the 200\,s run, with a median miss of 3\,MiB and a maximum miss of 82\,MiB. Thus, for the workloads studied, the initial 30\,s window provides both stable pressure indicators and an accurate current-memory snapshot for subsequent placement decisions.

\subsection{Threshold and Sequence Sensitivity}
\label{sec:eval:threshold_sensitivity}

After the activity-anchored observation window is complete, \sysname{} applies thresholds to the resulting pressure indicators to determine whether the GPU is saturated and should be withheld from additional collocation. These thresholds are operating points rather than hardware constants: the pressure indicators expose resource activity but do not define universal boundaries for beneficial sharing. We therefore evaluate sensitivity to threshold choice and workload arrival order.

We sweep threshold settings using calibrated one-GPU workload sequences containing small, medium, and pressure-heavy jobs in both \textbf{forward} and \textbf{reverse} arrival orders. Each arriving job is considered for collocation according to the pressure state produced by the workload(s) already running on the GPU. This setup stresses the saturation rule with memory-feasible placements that differ in compute and memory-system pressure and shows how placement order and overlap affect later decisions. We use these sequences only to select the default operating point; during deployment, \sysname{} derives pressure indicators online from runtime metric samples and applies the saturation rule without requiring submitted workloads or workload pairs to be profiled before scheduling. The calibrated sequences, complete sweep results, and analysis scripts are included in the evaluation artifacts.

\begin{figure}[t]
\centering
\includegraphics[width=\linewidth]{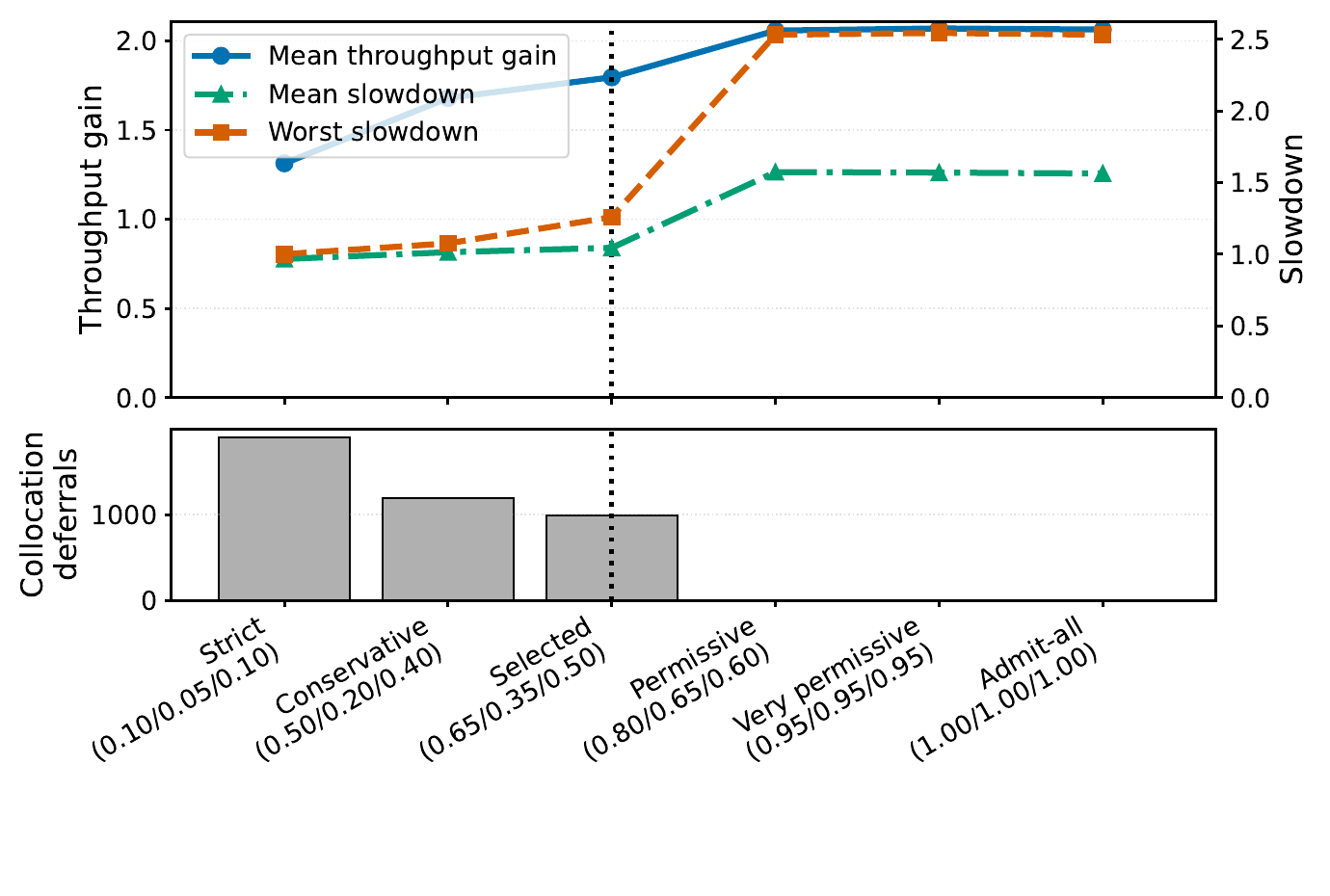}
\vspace{-10mm}
\caption{
Pressure-threshold trade-off on calibrated one-GPU workload sequences. The top panel shows throughput gain and slowdown; the bottom panel shows collocation deferrals. The selected operating point lies near the knee of the trade-off.
}
\label{fig:threshold_tradeoff}
\end{figure}

\Cref{fig:threshold_tradeoff} shows the throughput--slowdown trade-off across representative operating points. Strict thresholds defer more collocations, preserving isolation but leaving throughput untapped. More permissive thresholds expose additional sharing opportunities and improve throughput, but increase both mean and worst-case slowdown. The selected setting, $(\tau_s,\tau_o,\tau_d)=(0.65,0.35,0.50)$, lies near the knee of the trade-off: it achieves approximately $1.80\times$ average throughput gain while limiting the measured worst-case slowdown in the sweep to $1.26\times$.

The sequence-level results show why no threshold setting is universally optimal. Slowdown depends on the workloads present, the resources they stress, arrival order, and overlap duration. Pressure-heavy homogeneous sequences create the largest slowdown outliers, while heterogeneous sequences provide more complementary sharing opportunities. The sweep should therefore be interpreted as calibration evidence rather than a prediction of the exact slowdown distribution in the full-trace evaluation.

Overall, the selected thresholds avoid both the excessive deferrals of strict settings and the high worst-case slowdown of permissive settings. In the full traces, this operating point primarily improves makespan and queueing time, although some colocated jobs still experience longer execution times. This is the intended role of \sysname{}: controlled sharing improves multi-GPU server efficiency but does not guarantee low slowdown for every job.

\section{Discussion and Future Directions}
\label{sec:discussion}

\sysname{} shows that effective GPU collocation requires more than memory capacity: queue state, placement order, runtime pressure, memory behavior, and recovery jointly shape the outcome. Even measured peak memory determines only the memory-feasible candidate set; runtime interference, placement order, and overlap still shape execution-time and tail behavior. Offline workload characterization is also limited because runtime sharing behavior depends on workload mix, GPU generation, telemetry availability, placement order, and overlap duration. Pressure thresholds should therefore be treated as operating points rather than hardware constants.

Several extensions remain open. \sysname{} currently optimizes efficiency rather than fairness or progress-aware slowdown. Future work could combine pressure monitoring with progress signals such as iteration time or samples per second to trigger isolation, migration, or reassignment, and could adapt thresholds using observed pressure and scheduling outcomes.
\section{Conclusion}
\label{sec:conclusion}

This paper presented \sysname{}, a server-scale runtime scheduler for controlled collocation of deep learning training workloads. \sysname{} combines memory feasibility, mandatory activity-anchored post-placement observation, runtime-pressure filtering, configurable placement, and OOM-aware recovery to make collocation decisions using current GPU state rather than relying solely on offline profiling or prediction.

Across heterogeneous workload traces, AEGIS reduces makespan and GPU energy-to-solution relative to exclusive execution and representative baselines by increasing GPU concurrency and reducing queueing time. These gains expose the throughput--per-task-slowdown trade-off: some jobs execute longer under interference even as overall makespan and response time improve.

Overall, \sysname{} shows that controlled collocation need not choose between exhaustive offline pairwise characterization and interference-agnostic sharing. Practical server-scale GPU collocation therefore requires trustworthy runtime state, informed placement, and graceful recovery from memory-infeasible decisions.

\bibliographystyle{ACM-Reference-Format}
\bibliography{main}


\begin{thebibliography}{49}


\ifx \showCODEN    \undefined \def \showCODEN     #1{\unskip}     \fi
\ifx \showISBNx    \undefined \def \showISBNx     #1{\unskip}     \fi
\ifx \showISBNxiii \undefined \def \showISBNxiii  #1{\unskip}     \fi
\ifx \showISSN     \undefined \def \showISSN      #1{\unskip}     \fi
\ifx \showLCCN     \undefined \def \showLCCN      #1{\unskip}     \fi
\ifx \shownote     \undefined \def \shownote      #1{#1}          \fi
\ifx \showarticletitle \undefined \def \showarticletitle #1{#1}   \fi
\ifx \showURL      \undefined \def \showURL       {\relax}        \fi
\providecommand\bibfield[2]{#2}
\providecommand\bibinfo[2]{#2}
\providecommand\natexlab[1]{#1}
\providecommand\showeprint[2][]{arXiv:#2}

\bibitem[Bai et~al\mbox{.}(2020)]%
        {PipeSwitch}
\bibfield{author}{\bibinfo{person}{Zhihao Bai}, \bibinfo{person}{Zhen Zhang}, \bibinfo{person}{Yibo Zhu}, {and} \bibinfo{person}{Xin Jin}.} \bibinfo{year}{2020}\natexlab{}.
\newblock \showarticletitle{{PipeSwitch}: Fast Pipelined Context Switching for Deep Learning Applications}. In \bibinfo{booktitle}{\emph{14th USENIX Symposium on Operating Systems Design and Implementation (OSDI 20)}}. \bibinfo{publisher}{USENIX Association}, \bibinfo{pages}{499--514}.
\newblock
\showISBNx{978-1-939133-19-9}
\urldef\tempurl%
\url{https://www.usenix.org/conference/osdi20/presentation/bai}
\showURL{%
\tempurl}


\bibitem[Cao et~al\mbox{.}(2023)]%
        {CaoSIAK23}
\bibfield{author}{\bibinfo{person}{Jiashen Cao}, \bibinfo{person}{Rathijit Sen}, \bibinfo{person}{Matteo Interlandi}, \bibinfo{person}{Joy Arulraj}, {and} \bibinfo{person}{Hyesoon Kim}.} \bibinfo{year}{2023}\natexlab{}.
\newblock \showarticletitle{{GPU} Database Systems Characterization and Optimization}.
\newblock \bibinfo{journal}{\emph{Proc. {VLDB} Endow.}} \bibinfo{volume}{17}, \bibinfo{number}{3} (\bibinfo{year}{2023}), \bibinfo{pages}{441--454}.
\newblock
\href{https://doi.org/10.14778/3632093.3632107}{doi:\nolinkurl{10.14778/3632093.3632107}}


\bibitem[Chaudhary et~al\mbox{.}(2020)]%
        {gandiva-fair}
\bibfield{author}{\bibinfo{person}{Shubham Chaudhary}, \bibinfo{person}{Ramachandran Ramjee}, \bibinfo{person}{Muthian Sivathanu}, \bibinfo{person}{Nipun Kwatra}, {and} \bibinfo{person}{Srinidhi Viswanatha}.} \bibinfo{year}{2020}\natexlab{}.
\newblock \showarticletitle{Balancing efficiency and fairness in heterogeneous GPU clusters for deep learning}. In \bibinfo{booktitle}{\emph{Proceedings of the Fifteenth European Conference on Computer Systems}} (Heraklion, Greece) \emph{(\bibinfo{series}{EuroSys '20})}. \bibinfo{publisher}{Association for Computing Machinery}, \bibinfo{address}{New York, NY, USA}, Article \bibinfo{articleno}{1}, \bibinfo{numpages}{16}~pages.
\newblock
\showISBNx{9781450368827}
\href{https://doi.org/10.1145/3342195.3387555}{doi:\nolinkurl{10.1145/3342195.3387555}}


\bibitem[{Criteo}(2015)]%
        {criteo}
\bibfield{author}{\bibinfo{person}{{Criteo}}.} \bibinfo{year}{2015}\natexlab{}.
\newblock \bibinfo{title}{{Criteo 1TB Click Logs dataset}}.
\newblock \bibinfo{howpublished}{\url{https://ailab.criteo.com/download-criteo-1tb-click-logs-dataset/}}.
\newblock
\newblock
\shownote{Dataset. Accessed: 2026-07-10}.


\bibitem[Deng et~al\mbox{.}(2009)]%
        {imagenet}
\bibfield{author}{\bibinfo{person}{Jia Deng}, \bibinfo{person}{Wei Dong}, \bibinfo{person}{Richard Socher}, \bibinfo{person}{Li-Jia Li}, \bibinfo{person}{Kai Li}, {and} \bibinfo{person}{Li Fei-Fei}.} \bibinfo{year}{2009}\natexlab{}.
\newblock \showarticletitle{ImageNet: A large-scale hierarchical image database}. In \bibinfo{booktitle}{\emph{2009 IEEE Conference on Computer Vision and Pattern Recognition}}. \bibinfo{pages}{248--255}.
\newblock
\href{https://doi.org/10.1109/CVPR.2009.5206848}{doi:\nolinkurl{10.1109/CVPR.2009.5206848}}


\bibitem[Everingham et~al\mbox{.}(2010)]%
        {pascal-voc}
\bibfield{author}{\bibinfo{person}{Mark Everingham}, \bibinfo{person}{Luc Van~Gool}, \bibinfo{person}{Christopher K.~I. Williams}, \bibinfo{person}{John Winn}, {and} \bibinfo{person}{Andrew Zisserman}.} \bibinfo{year}{2010}\natexlab{}.
\newblock \showarticletitle{The Pascal Visual Object Classes (VOC) Challenge}.
\newblock \bibinfo{journal}{\emph{International Journal of Computer Vision}} \bibinfo{volume}{88}, \bibinfo{number}{2} (\bibinfo{year}{2010}), \bibinfo{pages}{303--338}.
\newblock
\href{https://doi.org/10.1007/s11263-009-0275-4}{doi:\nolinkurl{10.1007/s11263-009-0275-4}}


\bibitem[Gao et~al\mbox{.}(2022)]%
        {gao2022deep}
\bibfield{author}{\bibinfo{person}{Wei Gao}, \bibinfo{person}{Qinghao Hu}, \bibinfo{person}{Zhisheng Ye}, \bibinfo{person}{Peng Sun}, \bibinfo{person}{Xiaolin Wang}, \bibinfo{person}{Yingwei Luo}, \bibinfo{person}{Tianwei Zhang}, {and} \bibinfo{person}{Yonggang Wen}.} \bibinfo{year}{2022}\natexlab{}.
\newblock \showarticletitle{Deep learning workload scheduling in gpu datacenters: Taxonomy, challenges and vision}.
\newblock \bibinfo{journal}{\emph{arXiv preprint arXiv:2205.11913}} (\bibinfo{year}{2022}).
\newblock


\bibitem[Gao et~al\mbox{.}(2023)]%
        {gao2023runtime}
\bibfield{author}{\bibinfo{person}{Yanjie Gao}, \bibinfo{person}{Xianyu Gu}, \bibinfo{person}{Hongyu Zhang}, \bibinfo{person}{Haoxiang Lin}, {and} \bibinfo{person}{Mao Yang}.} \bibinfo{year}{2023}\natexlab{}.
\newblock \showarticletitle{Runtime Performance Prediction for Deep Learning Models with Graph Neural Network}. In \bibinfo{booktitle}{\emph{2023 IEEE/ACM 45th International Conference on Software Engineering: Software Engineering in Practice (ICSE-SEIP)}}. \bibinfo{pages}{368--380}.
\newblock
\href{https://doi.org/10.1109/ICSE-SEIP58684.2023.00039}{doi:\nolinkurl{10.1109/ICSE-SEIP58684.2023.00039}}


\bibitem[Gao et~al\mbox{.}(2020)]%
        {gao2020estimating}
\bibfield{author}{\bibinfo{person}{Yanjie Gao}, \bibinfo{person}{Yu Liu}, \bibinfo{person}{Hongyu Zhang}, \bibinfo{person}{Zhengxian Li}, \bibinfo{person}{Yonghao Zhu}, \bibinfo{person}{Haoxiang Lin}, {and} \bibinfo{person}{Mao Yang}.} \bibinfo{year}{2020}\natexlab{}.
\newblock \showarticletitle{Estimating GPU memory consumption of deep learning models}. In \bibinfo{booktitle}{\emph{Proceedings of the 28th ACM Joint Meeting on European Software Engineering Conference and Symposium on the Foundations of Software Engineering}} (Virtual Event, USA) \emph{(\bibinfo{series}{ESEC/FSE 2020})}. \bibinfo{publisher}{Association for Computing Machinery}, \bibinfo{address}{New York, NY, USA}, \bibinfo{pages}{1342–1352}.
\newblock
\showISBNx{9781450370431}
\href{https://doi.org/10.1145/3368089.3417050}{doi:\nolinkurl{10.1145/3368089.3417050}}


\bibitem[Gu et~al\mbox{.}(2023)]%
        {elasticFlow}
\bibfield{author}{\bibinfo{person}{Diandian Gu}, \bibinfo{person}{Yihao Zhao}, \bibinfo{person}{Yinmin Zhong}, \bibinfo{person}{Yifan Xiong}, \bibinfo{person}{Zhenhua Han}, \bibinfo{person}{Peng Cheng}, \bibinfo{person}{Fan Yang}, \bibinfo{person}{Gang Huang}, \bibinfo{person}{Xin Jin}, {and} \bibinfo{person}{Xuanzhe Liu}.} \bibinfo{year}{2023}\natexlab{}.
\newblock \showarticletitle{ElasticFlow: An Elastic Serverless Training Platform for Distributed Deep Learning}. In \bibinfo{booktitle}{\emph{Proceedings of the 28th ACM International Conference on Architectural Support for Programming Languages and Operating Systems, Volume 2}} (Vancouver, BC, Canada) \emph{(\bibinfo{series}{ASPLOS 2023})}. \bibinfo{publisher}{Association for Computing Machinery}, \bibinfo{address}{New York, NY, USA}, \bibinfo{pages}{266–280}.
\newblock
\showISBNx{9781450399166}
\href{https://doi.org/10.1145/3575693.3575721}{doi:\nolinkurl{10.1145/3575693.3575721}}


\bibitem[Gu et~al\mbox{.}(2019)]%
        {Tiresias}
\bibfield{author}{\bibinfo{person}{Juncheng Gu}, \bibinfo{person}{Mosharaf Chowdhury}, \bibinfo{person}{Kang~G. Shin}, \bibinfo{person}{Yibo Zhu}, \bibinfo{person}{Myeongjae Jeon}, \bibinfo{person}{Junjie Qian}, \bibinfo{person}{Hongqiang Liu}, {and} \bibinfo{person}{Chuanxiong Guo}.} \bibinfo{year}{2019}\natexlab{}.
\newblock \showarticletitle{Tiresias: A {GPU} Cluster Manager for Distributed Deep Learning}. In \bibinfo{booktitle}{\emph{16th USENIX Symposium on Networked Systems Design and Implementation (NSDI 19)}}. \bibinfo{publisher}{USENIX Association}, \bibinfo{address}{Boston, MA}, \bibinfo{pages}{485--500}.
\newblock
\showISBNx{978-1-931971-49-2}
\urldef\tempurl%
\url{https://www.usenix.org/conference/nsdi19/presentation/gu}
\showURL{%
\tempurl}


\bibitem[Harris(2015)]%
        {gpu_stream}
\bibfield{author}{\bibinfo{person}{Mark Harris}.} \bibinfo{year}{2015}\natexlab{}.
\newblock \bibinfo{title}{{GPU Pro Tip: CUDA 7 Streams Simplify Concurrency}}.
\newblock \bibinfo{howpublished}{\url{https://developer.nvidia.com/blog/gpu-pro-tip-cuda-7-streams-simplify-concurrency/}}.
\newblock
\newblock
\shownote{NVIDIA Technical Blog. Accessed: 2026-07-10}.


\bibitem[Hu et~al\mbox{.}(2021)]%
        {hu2021characterization}
\bibfield{author}{\bibinfo{person}{Qinghao Hu}, \bibinfo{person}{Peng Sun}, \bibinfo{person}{Shengen Yan}, \bibinfo{person}{Yonggang Wen}, {and} \bibinfo{person}{Tianwei Zhang}.} \bibinfo{year}{2021}\natexlab{}.
\newblock \showarticletitle{Characterization and prediction of deep learning workloads in large-scale GPU datacenters}. In \bibinfo{booktitle}{\emph{Proceedings of the International Conference for High Performance Computing, Networking, Storage and Analysis}} (St. Louis, Missouri) \emph{(\bibinfo{series}{SC '21})}. \bibinfo{publisher}{Association for Computing Machinery}, \bibinfo{address}{New York, NY, USA}, Article \bibinfo{articleno}{104}, \bibinfo{numpages}{15}~pages.
\newblock
\showISBNx{9781450384421}
\href{https://doi.org/10.1145/3458817.3476223}{doi:\nolinkurl{10.1145/3458817.3476223}}


\bibitem[Hu et~al\mbox{.}(2023)]%
        {lucid_asplos}
\bibfield{author}{\bibinfo{person}{Qinghao Hu}, \bibinfo{person}{Meng Zhang}, \bibinfo{person}{Peng Sun}, \bibinfo{person}{Yonggang Wen}, {and} \bibinfo{person}{Tianwei Zhang}.} \bibinfo{year}{2023}\natexlab{}.
\newblock \showarticletitle{Lucid: A Non-intrusive, Scalable and Interpretable Scheduler for Deep Learning Training Jobs}. In \bibinfo{booktitle}{\emph{Proceedings of the 28th ACM International Conference on Architectural Support for Programming Languages and Operating Systems, Volume 2}} (Vancouver, BC, Canada) \emph{(\bibinfo{series}{ASPLOS 2023})}. \bibinfo{publisher}{Association for Computing Machinery}, \bibinfo{address}{New York, NY, USA}, \bibinfo{pages}{457–472}.
\newblock
\showISBNx{9781450399166}
\href{https://doi.org/10.1145/3575693.3575705}{doi:\nolinkurl{10.1145/3575693.3575705}}


\bibitem[Hwang et~al\mbox{.}(2021)]%
        {AFS/CoDDL}
\bibfield{author}{\bibinfo{person}{Changho Hwang}, \bibinfo{person}{Taehyun Kim}, \bibinfo{person}{Sunghyun Kim}, \bibinfo{person}{Jinwoo Shin}, {and} \bibinfo{person}{KyoungSoo Park}.} \bibinfo{year}{2021}\natexlab{}.
\newblock \showarticletitle{Elastic Resource Sharing for Distributed Deep Learning}. In \bibinfo{booktitle}{\emph{18th USENIX Symposium on Networked Systems Design and Implementation (NSDI 21)}}. \bibinfo{publisher}{USENIX Association}, \bibinfo{pages}{721--739}.
\newblock
\showISBNx{978-1-939133-21-2}
\urldef\tempurl%
\url{https://www.usenix.org/conference/nsdi21/presentation/hwang}
\showURL{%
\tempurl}


\bibitem[Jayaram~Subramanya et~al\mbox{.}(2023)]%
        {sia}
\bibfield{author}{\bibinfo{person}{Suhas Jayaram~Subramanya}, \bibinfo{person}{Daiyaan Arfeen}, \bibinfo{person}{Shouxu Lin}, \bibinfo{person}{Aurick Qiao}, \bibinfo{person}{Zhihao Jia}, {and} \bibinfo{person}{Gregory~R. Ganger}.} \bibinfo{year}{2023}\natexlab{}.
\newblock \showarticletitle{Sia: Heterogeneity-aware, goodput-optimized ML-cluster scheduling}. In \bibinfo{booktitle}{\emph{Proceedings of the 29th Symposium on Operating Systems Principles}} (Koblenz, Germany) \emph{(\bibinfo{series}{SOSP '23})}. \bibinfo{publisher}{Association for Computing Machinery}, \bibinfo{address}{New York, NY, USA}, \bibinfo{pages}{642–657}.
\newblock
\showISBNx{9798400702297}
\href{https://doi.org/10.1145/3600006.3613175}{doi:\nolinkurl{10.1145/3600006.3613175}}


\bibitem[Jeon et~al\mbox{.}(2019)]%
        {AnalysisOfMultiTenantGPUClusters}
\bibfield{author}{\bibinfo{person}{Myeongjae Jeon}, \bibinfo{person}{Shivaram Venkataraman}, \bibinfo{person}{Amar Phanishayee}, \bibinfo{person}{Junjie Qian}, \bibinfo{person}{Wencong Xiao}, {and} \bibinfo{person}{Fan Yang}.} \bibinfo{year}{2019}\natexlab{}.
\newblock \showarticletitle{Analysis of {Large-Scale} {Multi-Tenant} {GPU} Clusters for {DNN} Training Workloads}. In \bibinfo{booktitle}{\emph{2019 USENIX Annual Technical Conference (USENIX ATC 19)}}. \bibinfo{publisher}{USENIX Association}, \bibinfo{address}{Renton, WA}, \bibinfo{pages}{947--960}.
\newblock
\showISBNx{978-1-939133-03-8}
\urldef\tempurl%
\url{https://www.usenix.org/conference/atc19/presentation/jeon}
\showURL{%
\tempurl}


\bibitem[Kim et~al\mbox{.}(2024)]%
        {kim2024llmem}
\bibfield{author}{\bibinfo{person}{Taeho Kim}, \bibinfo{person}{Yanming Wang}, \bibinfo{person}{Vatshank Chaturvedi}, \bibinfo{person}{Lokesh Gupta}, \bibinfo{person}{Seyeon Kim}, \bibinfo{person}{Yongin Kwon}, {and} \bibinfo{person}{Sangtae Ha}.} \bibinfo{year}{2024}\natexlab{}.
\newblock \showarticletitle{LLMem: estimating GPU memory usage for fine-tuning pre-trained LLMs}. In \bibinfo{booktitle}{\emph{Proceedings of the Thirty-Third International Joint Conference on Artificial Intelligence}} (Jeju, Korea) \emph{(\bibinfo{series}{IJCAI '24})}. Article \bibinfo{articleno}{699}, \bibinfo{numpages}{9}~pages.
\newblock
\showISBNx{978-1-956792-04-1}
\href{https://doi.org/10.24963/ijcai.2024/699}{doi:\nolinkurl{10.24963/ijcai.2024/699}}


\bibitem[Krizhevsky(2009)]%
        {cifar}
\bibfield{author}{\bibinfo{person}{Alex Krizhevsky}.} \bibinfo{year}{2009}\natexlab{}.
\newblock \bibinfo{booktitle}{\emph{Learning Multiple Layers of Features from Tiny Images}}.
\newblock \bibinfo{type}{{T}echnical {R}eport}. \bibinfo{institution}{University of Toronto}.
\newblock


\bibitem[Li et~al\mbox{.}(2023)]%
        {lyra}
\bibfield{author}{\bibinfo{person}{Jiamin Li}, \bibinfo{person}{Hong Xu}, \bibinfo{person}{Yibo Zhu}, \bibinfo{person}{Zherui Liu}, \bibinfo{person}{Chuanxiong Guo}, {and} \bibinfo{person}{Cong Wang}.} \bibinfo{year}{2023}\natexlab{}.
\newblock \showarticletitle{Lyra: Elastic Scheduling for Deep Learning Clusters}. In \bibinfo{booktitle}{\emph{Proceedings of the Eighteenth European Conference on Computer Systems}} (Rome, Italy) \emph{(\bibinfo{series}{EuroSys '23})}. \bibinfo{publisher}{Association for Computing Machinery}, \bibinfo{address}{New York, NY, USA}, \bibinfo{pages}{835–850}.
\newblock
\showISBNx{9781450394871}
\href{https://doi.org/10.1145/3552326.3587445}{doi:\nolinkurl{10.1145/3552326.3587445}}


\bibitem[Lin et~al\mbox{.}(2014)]%
        {coco}
\bibfield{author}{\bibinfo{person}{Tsung-Yi Lin}, \bibinfo{person}{Michael Maire}, \bibinfo{person}{Serge Belongie}, \bibinfo{person}{James Hays}, \bibinfo{person}{Pietro Perona}, \bibinfo{person}{Deva Ramanan}, \bibinfo{person}{Piotr Doll{\'a}r}, {and} \bibinfo{person}{C.~Lawrence Zitnick}.} \bibinfo{year}{2014}\natexlab{}.
\newblock \showarticletitle{Microsoft COCO: Common Objects in Context}. In \bibinfo{booktitle}{\emph{European Conference on Computer Vision (ECCV)}}. \bibinfo{publisher}{Springer}, \bibinfo{pages}{740--755}.
\newblock
\href{https://doi.org/10.1007/978-3-319-10602-1_48}{doi:\nolinkurl{10.1007/978-3-319-10602-1_48}}


\bibitem[Mahajan et~al\mbox{.}(2020)]%
        {Themis}
\bibfield{author}{\bibinfo{person}{Kshiteej Mahajan}, \bibinfo{person}{Arjun Balasubramanian}, \bibinfo{person}{Arjun Singhvi}, \bibinfo{person}{Shivaram Venkataraman}, \bibinfo{person}{Aditya Akella}, \bibinfo{person}{Amar Phanishayee}, {and} \bibinfo{person}{Shuchi Chawla}.} \bibinfo{year}{2020}\natexlab{}.
\newblock \showarticletitle{Themis: Fair and Efficient {GPU} Cluster Scheduling}. In \bibinfo{booktitle}{\emph{17th USENIX Symposium on Networked Systems Design and Implementation (NSDI 20)}}. \bibinfo{publisher}{USENIX Association}, \bibinfo{address}{Santa Clara, CA}, \bibinfo{pages}{289--304}.
\newblock
\showISBNx{978-1-939133-13-7}
\urldef\tempurl%
\url{https://www.usenix.org/conference/nsdi20/presentation/mahajan}
\showURL{%
\tempurl}


\bibitem[Mai et~al\mbox{.}(2020)]%
        {KungFu}
\bibfield{author}{\bibinfo{person}{Luo Mai}, \bibinfo{person}{Guo Li}, \bibinfo{person}{Marcel Wagenl{\"a}nder}, \bibinfo{person}{Konstantinos Fertakis}, \bibinfo{person}{Andrei-Octavian Brabete}, {and} \bibinfo{person}{Peter Pietzuch}.} \bibinfo{year}{2020}\natexlab{}.
\newblock \showarticletitle{{KungFu}: Making Training in Distributed Machine Learning Adaptive}. In \bibinfo{booktitle}{\emph{14th USENIX Symposium on Operating Systems Design and Implementation (OSDI 20)}}. \bibinfo{publisher}{USENIX Association}, \bibinfo{pages}{937--954}.
\newblock
\showISBNx{978-1-939133-19-9}
\urldef\tempurl%
\url{https://www.usenix.org/conference/osdi20/presentation/mai}
\showURL{%
\tempurl}


\bibitem[{Memory Requirements}(2025)]%
        {deepspeed_memory_requirements}
\bibfield{author}{\bibinfo{person}{{Memory Requirements}}.} \bibinfo{year}{2025}\natexlab{}.
\newblock \bibinfo{title}{{DeepSpeed}}.
\newblock \bibinfo{howpublished}{\url{https://deepspeed.readthedocs.io/en/latest/memory.html}}.
\newblock
\newblock
\shownote{DeepSpeed documentation. Accessed: 2026-07-10}.


\bibitem[Merity et~al\mbox{.}(2017)]%
        {wikitext}
\bibfield{author}{\bibinfo{person}{Stephen Merity}, \bibinfo{person}{Caiming Xiong}, \bibinfo{person}{James Bradbury}, {and} \bibinfo{person}{Richard Socher}.} \bibinfo{year}{2017}\natexlab{}.
\newblock \showarticletitle{Pointer Sentinel Mixture Models}. In \bibinfo{booktitle}{\emph{International Conference on Learning Representations}}.
\newblock
\urldef\tempurl%
\url{https://openreview.net/forum?id=Byj72udxe}
\showURL{%
\tempurl}


\bibitem[{Microsoft Research}(2019)]%
        {philly_traces}
\bibfield{author}{\bibinfo{person}{{Microsoft Research}}.} \bibinfo{year}{2019}\natexlab{}.
\newblock \bibinfo{title}{{Philly GPU Cluster Trace}}.
\newblock \bibinfo{howpublished}{\url{https://github.com/msr-fiddle/philly-traces}}.
\newblock
\newblock
\shownote{Dataset. Accessed: 2026-07-10}.


\bibitem[Narayanan et~al\mbox{.}(2020)]%
        {gavel}
\bibfield{author}{\bibinfo{person}{Deepak Narayanan}, \bibinfo{person}{Keshav Santhanam}, \bibinfo{person}{Fiodar Kazhamiaka}, \bibinfo{person}{Amar Phanishayee}, {and} \bibinfo{person}{Matei Zaharia}.} \bibinfo{year}{2020}\natexlab{}.
\newblock \showarticletitle{{Heterogeneity-Aware} Cluster Scheduling Policies for Deep Learning Workloads}. In \bibinfo{booktitle}{\emph{14th USENIX Symposium on Operating Systems Design and Implementation (OSDI 20)}}. \bibinfo{publisher}{USENIX Association}, \bibinfo{pages}{481--498}.
\newblock
\showISBNx{978-1-939133-19-9}
\urldef\tempurl%
\url{https://www.usenix.org/conference/osdi20/presentation/narayanan-deepak}
\showURL{%
\tempurl}


\bibitem[{NVIDIA Corporation}(2025a)]%
        {nvidia_mps}
\bibfield{author}{\bibinfo{person}{{NVIDIA Corporation}}.} \bibinfo{year}{2025}\natexlab{a}.
\newblock \bibinfo{booktitle}{\emph{{Multi-Process Service}}}.
\newblock
\urldef\tempurl%
\url{https://docs.nvidia.com/deploy/mps/index.html}
\showURL{%
\tempurl}
\newblock
\shownote{NVIDIA GPU Management and Deployment Documentation. Accessed: 2026-07-10}.


\bibitem[{NVIDIA Corporation}(2025b)]%
        {mig-main-guide-pdf}
\bibfield{author}{\bibinfo{person}{{NVIDIA Corporation}}.} \bibinfo{year}{2025}\natexlab{b}.
\newblock \bibinfo{booktitle}{\emph{{NVIDIA Multi-Instance GPU User Guide}}}.
\newblock
\urldef\tempurl%
\url{https://docs.nvidia.com/datacenter/tesla/pdf/MIG_User_Guide.pdf}
\showURL{%
\tempurl}
\newblock
\shownote{Accessed: 2026-07-10}.


\bibitem[{NVIDIA Corporation}(nda)]%
        {dcgm}
\bibfield{author}{\bibinfo{person}{{NVIDIA Corporation}}.} \bibinfo{year}{{n.d.}}\natexlab{a}.
\newblock \bibinfo{title}{{NVIDIA Data Center GPU Manager}}.
\newblock \bibinfo{howpublished}{\url{https://github.com/NVIDIA/DCGM}}.
\newblock
\newblock
\shownote{GitHub repository. Accessed: 2026-07-10}.


\bibitem[{NVIDIA Corporation}(ndb)]%
        {dcgm-field-ids}
\bibfield{author}{\bibinfo{person}{{NVIDIA Corporation}}.} \bibinfo{year}{{n.d.}}\natexlab{b}.
\newblock \bibinfo{title}{{NVIDIA Data Center GPU Manager 3.0: Field Identifiers}}.
\newblock \bibinfo{howpublished}{\url{https://docs.nvidia.com/datacenter/dcgm/3.0/dcgm-api/dcgm-api-field-ids.html}}.
\newblock
\newblock
\shownote{NVIDIA DCGM documentation. Accessed: 2026-07-10}.


\bibitem[Peng et~al\mbox{.}(2018)]%
        {optimus}
\bibfield{author}{\bibinfo{person}{Yanghua Peng}, \bibinfo{person}{Yixin Bao}, \bibinfo{person}{Yangrui Chen}, \bibinfo{person}{Chuan Wu}, {and} \bibinfo{person}{Chuanxiong Guo}.} \bibinfo{year}{2018}\natexlab{}.
\newblock \showarticletitle{Optimus: an efficient dynamic resource scheduler for deep learning clusters}. In \bibinfo{booktitle}{\emph{Proceedings of the Thirteenth EuroSys Conference}} (Porto, Portugal) \emph{(\bibinfo{series}{EuroSys '18})}. \bibinfo{publisher}{Association for Computing Machinery}, \bibinfo{address}{New York, NY, USA}, Article \bibinfo{articleno}{3}, \bibinfo{numpages}{14}~pages.
\newblock
\showISBNx{9781450355841}
\href{https://doi.org/10.1145/3190508.3190517}{doi:\nolinkurl{10.1145/3190508.3190517}}


\bibitem[{PyTorch Contributors}(nd)]%
        {faketensor2023}
\bibfield{author}{\bibinfo{person}{{PyTorch Contributors}}.} \bibinfo{year}{{n.d.}}\natexlab{}.
\newblock \bibinfo{title}{{Fake Tensor PyTorch 2.13 Documentation}}.
\newblock \bibinfo{howpublished}{\url{https://pytorch.org/docs/stable/torch.compiler_fake_tensor.html}}.
\newblock
\newblock
\shownote{PyTorch documentation. Accessed: 2026-07-10}.


\bibitem[Qiao et~al\mbox{.}(2021)]%
        {Pollux}
\bibfield{author}{\bibinfo{person}{Aurick Qiao}, \bibinfo{person}{Sang~Keun Choe}, \bibinfo{person}{Suhas~Jayaram Subramanya}, \bibinfo{person}{Willie Neiswanger}, \bibinfo{person}{Qirong Ho}, \bibinfo{person}{Hao Zhang}, \bibinfo{person}{Gregory~R. Ganger}, {and} \bibinfo{person}{Eric~P. Xing}.} \bibinfo{year}{2021}\natexlab{}.
\newblock \showarticletitle{Pollux: Co-adaptive Cluster Scheduling for Goodput-Optimized Deep Learning}. In \bibinfo{booktitle}{\emph{15th {USENIX} Symposium on Operating Systems Design and Implementation ({OSDI} 21)}}. \bibinfo{publisher}{{USENIX} Association}, \bibinfo{pages}{1--18}.
\newblock
\showISBNx{978-1-939133-22-9}
\urldef\tempurl%
\url{https://www.usenix.org/conference/osdi21/presentation/qiao}
\showURL{%
\tempurl}


\bibitem[Robroek et~al\mbox{.}(2024)]%
        {robroek2023analysis}
\bibfield{author}{\bibinfo{person}{Ties Robroek}, \bibinfo{person}{Ehsan Yousefzadeh-Asl-Miandoab}, {and} \bibinfo{person}{P\i{}nar T\"{o}z\"{u}n}.} \bibinfo{year}{2024}\natexlab{}.
\newblock \showarticletitle{An Analysis of Collocation on GPUs for Deep Learning Training}. In \bibinfo{booktitle}{\emph{Proceedings of the 4th Workshop on Machine Learning and Systems}} (Athens, Greece) \emph{(\bibinfo{series}{EuroMLSys '24})}. \bibinfo{publisher}{Association for Computing Machinery}, \bibinfo{address}{New York, NY, USA}, \bibinfo{pages}{81–90}.
\newblock
\showISBNx{9798400705410}
\href{https://doi.org/10.1145/3642970.3655827}{doi:\nolinkurl{10.1145/3642970.3655827}}


\bibitem[{S-Lab System Group}(2021)]%
        {heliosdata}
\bibfield{author}{\bibinfo{person}{{S-Lab System Group}}.} \bibinfo{year}{2021}\natexlab{}.
\newblock \bibinfo{title}{{HeliosData}: Helios Traces from {SenseTime}}.
\newblock \bibinfo{howpublished}{GitHub repository, HeliosData release}.
\newblock
\urldef\tempurl%
\url{https://github.com/S-Lab-System-Group/HeliosData}
\showURL{%
\tempurl}
\newblock
\shownote{Dataset. Released July 19, 2021. Accessed: 2026-07-10}.


\bibitem[Shi et~al\mbox{.}(2025)]%
        {10.1145/3721462.3770773}
\bibfield{author}{\bibinfo{person}{Jiabo Shi}, \bibinfo{person}{Dimitrios Pezaros}, {and} \bibinfo{person}{Yehia Elkhatib}.} \bibinfo{year}{2025}\natexlab{}.
\newblock \showarticletitle{xMem: A CPU-Based Approach for Accurate Estimation of GPU Memory in Deep Learning Training Workloads}. In \bibinfo{booktitle}{\emph{Proceedings of the 26th International Middleware Conference}} (Vanderbilt University, Nashville, TN, USA) \emph{(\bibinfo{series}{Middleware '25})}. \bibinfo{publisher}{Association for Computing Machinery}, \bibinfo{address}{New York, NY, USA}, \bibinfo{pages}{256–269}.
\newblock
\showISBNx{9798400715549}
\href{https://doi.org/10.1145/3721462.3770773}{doi:\nolinkurl{10.1145/3721462.3770773}}


\bibitem[Strati et~al\mbox{.}(2024)]%
        {strati2024orion}
\bibfield{author}{\bibinfo{person}{Foteini Strati}, \bibinfo{person}{Xianzhe Ma}, {and} \bibinfo{person}{Ana Klimovic}.} \bibinfo{year}{2024}\natexlab{}.
\newblock \showarticletitle{Orion: Interference-aware, Fine-grained GPU Sharing for ML Applications}. In \bibinfo{booktitle}{\emph{Proceedings of the Nineteenth European Conference on Computer Systems}} (Athens, Greece) \emph{(\bibinfo{series}{EuroSys '24})}. \bibinfo{publisher}{Association for Computing Machinery}, \bibinfo{address}{New York, NY, USA}, \bibinfo{pages}{1075–1092}.
\newblock
\showISBNx{9798400704376}
\href{https://doi.org/10.1145/3627703.3629578}{doi:\nolinkurl{10.1145/3627703.3629578}}


\bibitem[Weng et~al\mbox{.}(2022)]%
        {MLaaS}
\bibfield{author}{\bibinfo{person}{Qizhen Weng}, \bibinfo{person}{Wencong Xiao}, \bibinfo{person}{Yinghao Yu}, \bibinfo{person}{Wei Wang}, \bibinfo{person}{Cheng Wang}, \bibinfo{person}{Jian He}, \bibinfo{person}{Yong Li}, \bibinfo{person}{Liping Zhang}, \bibinfo{person}{Wei Lin}, {and} \bibinfo{person}{Yu Ding}.} \bibinfo{year}{2022}\natexlab{}.
\newblock \showarticletitle{{MLaaS} in the Wild: Workload Analysis and Scheduling in {Large-Scale} Heterogeneous {GPU} Clusters}. In \bibinfo{booktitle}{\emph{19th USENIX Symposium on Networked Systems Design and Implementation (NSDI 22)}}. \bibinfo{publisher}{USENIX Association}, \bibinfo{address}{Renton, WA}, \bibinfo{pages}{945--960}.
\newblock
\showISBNx{978-1-939133-27-4}
\urldef\tempurl%
\url{https://www.usenix.org/conference/nsdi22/presentation/weng}
\showURL{%
\tempurl}


\bibitem[Xiao et~al\mbox{.}(2018)]%
        {Gandiva}
\bibfield{author}{\bibinfo{person}{Wencong Xiao}, \bibinfo{person}{Romil Bhardwaj}, \bibinfo{person}{Ramachandran Ramjee}, \bibinfo{person}{Muthian Sivathanu}, \bibinfo{person}{Nipun Kwatra}, \bibinfo{person}{Zhenhua Han}, \bibinfo{person}{Pratyush Patel}, \bibinfo{person}{Xuan Peng}, \bibinfo{person}{Hanyu Zhao}, \bibinfo{person}{Quanlu Zhang}, \bibinfo{person}{Fan Yang}, {and} \bibinfo{person}{Lidong Zhou}.} \bibinfo{year}{2018}\natexlab{}.
\newblock \showarticletitle{Gandiva: Introspective Cluster Scheduling for Deep Learning}. In \bibinfo{booktitle}{\emph{13th USENIX Symposium on Operating Systems Design and Implementation (OSDI 18)}}. \bibinfo{publisher}{USENIX Association}, \bibinfo{address}{Carlsbad, CA}, \bibinfo{pages}{595--610}.
\newblock
\showISBNx{978-1-939133-08-3}
\urldef\tempurl%
\url{https://www.usenix.org/conference/osdi18/presentation/xiao}
\showURL{%
\tempurl}


\bibitem[Xiao et~al\mbox{.}(2020)]%
        {AntMan}
\bibfield{author}{\bibinfo{person}{Wencong Xiao}, \bibinfo{person}{Shiru Ren}, \bibinfo{person}{Yong Li}, \bibinfo{person}{Yang Zhang}, \bibinfo{person}{Pengyang Hou}, \bibinfo{person}{Zhi Li}, \bibinfo{person}{Yihui Feng}, \bibinfo{person}{Wei Lin}, {and} \bibinfo{person}{Yangqing Jia}.} \bibinfo{year}{2020}\natexlab{}.
\newblock \showarticletitle{{AntMan}: Dynamic Scaling on {GPU} Clusters for Deep Learning}. In \bibinfo{booktitle}{\emph{14th USENIX Symposium on Operating Systems Design and Implementation (OSDI 20)}}. \bibinfo{publisher}{USENIX Association}, \bibinfo{pages}{533--548}.
\newblock
\showISBNx{978-1-939133-19-9}
\urldef\tempurl%
\url{https://www.usenix.org/conference/osdi20/presentation/xiao}
\showURL{%
\tempurl}


\bibitem[Ye et~al\mbox{.}(2024)]%
        {10.1145/3638757}
\bibfield{author}{\bibinfo{person}{Zhisheng Ye}, \bibinfo{person}{Wei Gao}, \bibinfo{person}{Qinghao Hu}, \bibinfo{person}{Peng Sun}, \bibinfo{person}{Xiaolin Wang}, \bibinfo{person}{Yingwei Luo}, \bibinfo{person}{Tianwei Zhang}, {and} \bibinfo{person}{Yonggang Wen}.} \bibinfo{year}{2024}\natexlab{}.
\newblock \showarticletitle{Deep Learning Workload Scheduling in GPU Datacenters: A Survey}.
\newblock \bibinfo{journal}{\emph{ACM Comput. Surv.}} \bibinfo{volume}{56}, \bibinfo{number}{6}, Article \bibinfo{articleno}{146} (\bibinfo{date}{Jan.} \bibinfo{year}{2024}), \bibinfo{numpages}{38}~pages.
\newblock
\showISSN{0360-0300}
\href{https://doi.org/10.1145/3638757}{doi:\nolinkurl{10.1145/3638757}}


\bibitem[Yeung et~al\mbox{.}(2022)]%
        {horus}
\bibfield{author}{\bibinfo{person}{Gingfung Yeung}, \bibinfo{person}{Damian Borowiec}, \bibinfo{person}{Renyu Yang}, \bibinfo{person}{Adrian Friday}, \bibinfo{person}{Richard Harper}, {and} \bibinfo{person}{Peter Garraghan}.} \bibinfo{year}{2022}\natexlab{}.
\newblock \showarticletitle{Horus: Interference-Aware and Prediction-Based Scheduling in Deep Learning Systems}.
\newblock \bibinfo{journal}{\emph{IEEE Transactions on Parallel and Distributed Systems}} \bibinfo{volume}{33}, \bibinfo{number}{1} (\bibinfo{year}{2022}), \bibinfo{pages}{88--100}.
\newblock
\href{https://doi.org/10.1109/TPDS.2021.3079202}{doi:\nolinkurl{10.1109/TPDS.2021.3079202}}


\bibitem[Yousefzadeh-Asl-Miandoab et~al\mbox{.}(2026)]%
        {yousefzadehGPUMemoryEstimation}
\bibfield{author}{\bibinfo{person}{Ehsan Yousefzadeh-Asl-Miandoab}, \bibinfo{person}{Reza Karimzadeh}, \bibinfo{person}{Danyal Yorulmaz}, \bibinfo{person}{Bulat Ibragimov}, {and} \bibinfo{person}{P\i{}nar T\"{o}z\"{u}n}.} \bibinfo{year}{2026}\natexlab{}.
\newblock \showarticletitle{GPU Memory and Utilization Estimation for Training-Aware Resource Management: Opportunities and Limitations}. In \bibinfo{booktitle}{\emph{Proceedings of the Sixth European Workshop on Machine Learning and Systems}} \emph{(\bibinfo{series}{EuroMLSys '26})}. \bibinfo{publisher}{Association for Computing Machinery}, \bibinfo{address}{New York, NY, USA}, \bibinfo{pages}{127–138}.
\newblock
\showISBNx{9798400726057}
\href{https://doi.org/10.1145/3805621.3807621}{doi:\nolinkurl{10.1145/3805621.3807621}}


\bibitem[Yousefzadeh-Asl-Miandoab et~al\mbox{.}(2023)]%
        {yousefzadeh2023profiling}
\bibfield{author}{\bibinfo{person}{Ehsan Yousefzadeh-Asl-Miandoab}, \bibinfo{person}{Ties Robroek}, {and} \bibinfo{person}{P\i{}nar T{\"{o}}z{\"{u}}n}.} \bibinfo{year}{2023}\natexlab{}.
\newblock \showarticletitle{Profiling and Monitoring Deep Learning Training Tasks}. In \bibinfo{booktitle}{\emph{Proceedings of the 3rd Workshop on Machine Learning and Systems}} (Rome, Italy) \emph{(\bibinfo{series}{EuroMLSys '23})}. \bibinfo{publisher}{Association for Computing Machinery}, \bibinfo{address}{New York, NY, USA}, \bibinfo{pages}{18–25}.
\newblock
\showISBNx{9798400700842}
\href{https://doi.org/10.1145/3578356.3592589}{doi:\nolinkurl{10.1145/3578356.3592589}}


\bibitem[Yu and Chowdhury(2020)]%
        {SalusFineGrainGPUSharingDL}
\bibfield{author}{\bibinfo{person}{Peifeng Yu} {and} \bibinfo{person}{Mosharaf Chowdhury}.} \bibinfo{year}{2020}\natexlab{}.
\newblock \showarticletitle{Fine-Grained GPU Sharing Primitives for Deep Learning Applications}. In \bibinfo{booktitle}{\emph{Proceedings of Machine Learning and Systems}}, \bibfield{editor}{\bibinfo{person}{I.~Dhillon}, \bibinfo{person}{D.~Papailiopoulos}, {and} \bibinfo{person}{V.~Sze}} (Eds.), Vol.~\bibinfo{volume}{2}. \bibinfo{pages}{98--111}.
\newblock
\urldef\tempurl%
\url{https://proceedings.mlsys.org/paper_files/paper/2020/file/d9cd83bc91b8c36a0c7c0fcca59228f2-Paper.pdf}
\showURL{%
\tempurl}


\bibitem[Zhao et~al\mbox{.}(2020)]%
        {HiveD}
\bibfield{author}{\bibinfo{person}{Hanyu Zhao}, \bibinfo{person}{Zhenhua Han}, \bibinfo{person}{Zhi Yang}, \bibinfo{person}{Quanlu Zhang}, \bibinfo{person}{Fan Yang}, \bibinfo{person}{Lidong Zhou}, \bibinfo{person}{Mao Yang}, \bibinfo{person}{Francis~C.M. Lau}, \bibinfo{person}{Yuqi Wang}, \bibinfo{person}{Yifan Xiong}, {and} \bibinfo{person}{Bin Wang}.} \bibinfo{year}{2020}\natexlab{}.
\newblock \showarticletitle{{HiveD}: Sharing a {GPU} Cluster for Deep Learning with Guarantees}. In \bibinfo{booktitle}{\emph{14th USENIX Symposium on Operating Systems Design and Implementation (OSDI 20)}}. \bibinfo{publisher}{USENIX Association}, \bibinfo{pages}{515--532}.
\newblock
\showISBNx{978-1-939133-19-9}
\urldef\tempurl%
\url{https://www.usenix.org/conference/osdi20/presentation/zhao-hanyu}
\showURL{%
\tempurl}


\bibitem[Zhao et~al\mbox{.}(2022)]%
        {Muri}
\bibfield{author}{\bibinfo{person}{Yihao Zhao}, \bibinfo{person}{Yuanqiang Liu}, \bibinfo{person}{Yanghua Peng}, \bibinfo{person}{Yibo Zhu}, \bibinfo{person}{Xuanzhe Liu}, {and} \bibinfo{person}{Xin Jin}.} \bibinfo{year}{2022}\natexlab{}.
\newblock \showarticletitle{Multi-resource interleaving for deep learning training}. In \bibinfo{booktitle}{\emph{Proceedings of the ACM SIGCOMM 2022 Conference}} (Amsterdam, Netherlands) \emph{(\bibinfo{series}{SIGCOMM '22})}. \bibinfo{publisher}{Association for Computing Machinery}, \bibinfo{address}{New York, NY, USA}, \bibinfo{pages}{428–440}.
\newblock
\showISBNx{9781450394208}
\href{https://doi.org/10.1145/3544216.3544224}{doi:\nolinkurl{10.1145/3544216.3544224}}


\bibitem[Zheng et~al\mbox{.}(2023)]%
        {shockWave}
\bibfield{author}{\bibinfo{person}{Pengfei Zheng}, \bibinfo{person}{Rui Pan}, \bibinfo{person}{Tarannum Khan}, \bibinfo{person}{Shivaram Venkataraman}, {and} \bibinfo{person}{Aditya Akella}.} \bibinfo{year}{2023}\natexlab{}.
\newblock \showarticletitle{Shockwave: Fair and Efficient Cluster Scheduling for Dynamic Adaptation in Machine Learning}. In \bibinfo{booktitle}{\emph{20th USENIX Symposium on Networked Systems Design and Implementation (NSDI 23)}}. \bibinfo{publisher}{USENIX Association}, \bibinfo{address}{Boston, MA}, \bibinfo{pages}{703--723}.
\newblock
\showISBNx{978-1-939133-33-5}
\urldef\tempurl%
\url{https://www.usenix.org/conference/nsdi23/presentation/zheng}
\showURL{%
\tempurl}


\end{thebibliography}

\clearpage

\appendix

\end{document}